\documentstyle[12pt]{article}
\begin{document}
\newcommand{\be}{\begin{equation}}
\newcommand{\ee}{\end{equation}}
\newcommand{\ba}{\begin{array}}
\newcommand{\ea}{\end{array}}
\newcommand{\bea}{\begin{eqnarray}}
\newcommand{\eea}{\end{eqnarray}}

\title{\bf Coulomb-oscillator duality in spaces of constant curvature }
\author{E.\ G.\ Kalnins \\
{\sl Department of Mathematics and Statistics,}\\
{\sl University of Waikato,}\\
{\sl Hamilton, New Zealand.}\\
W.\ Miller, Jr. \\
{\sl School of Mathematics, University of Minnesota,}\\
{\sl Minneapolis, Minnesota, 55455, U.S.A.}\\
and G.\ S.\ Pogosyan\thanks{%
Work supported in part by the Russian Foundation for Basic Research
under grant 98-01-00330\hfill\hfill} \\
{\sl Laboratory of Theoretical Physics,}\\
{\sl Joint Institute for Nuclear Research,}\\
{\sl Dubna, Moscow Region 141980, Russia}}
\date{\today}
\maketitle

{\noindent{\bf PACS}: 02.20.+b, 03.65.Fd}

\vspace{0.6cm}

\begin{center}
\noindent{\large {\bf Abstract}}
\end{center}

\vspace{0.3cm}

\noindent
In this paper we construct generalizations to spheres of the well
known Levi-Civita, Kustaanheimo-Steifel and Hurwitz regularizing
transformations in Euclidean spaces of dimensions 2, 3 and 5.
The corresponding classical and quantum mechanical analogues of
the Kepler-Coulomb problem on these spheres are  discussed.

\newpage

\section{Introduction}

It is well known that the problem of a body moving under the influence
of a central force field with potential $V(r)=-\mu /r$ has a singularity
at the origin. We refer to this as the {\it Kepler problem}.
This problem is usually posed in 3 dimensions, but since the motion
is always constrained to a plane perpendicular to the constant
angular momentum vector we can reduce it to 2 dimensions with
Newtonian equations of motion and energy integral
%=================================================================
\be
{d^2\over dt^2}\mbox{\boldmath{$ r$}}
=
- \frac{\mu}{r^3}\mbox{\boldmath{$ r$}}, \quad
{1\over 2}\left({dr\over dt}\right)^2
-{\mu \over r}+{1\over 2r^2} = h,
\label{coulomb}
\ee
%=================================================================
where $r^2=\mbox{\boldmath{$ r$}}\cdot\mbox{\boldmath{$ r$}}$,
$r^2{d\theta \over dt}=c$ and $\mbox{\boldmath{$ r$}}=(x,y)
=(r\cos\theta,r\sin\theta)$. As is well known \cite{LC,KS},
in two dimensions the Levi-Civita transformation effectively removes
the singularity and rewrites this problem in terms of the classical
harmonic oscillator. In this process the original problem has been
regularized. To achieve the regularization, instead of $t$ we use
the variable $s$ defined by
%================================================================
\be
s = \int \frac{dt}{r}, \qquad
{d\over dt} = {1\over r} \frac{d}{d s}.
\ee
%=================================================================
With $x'={dx\over ds}$ etc., the original equations (\ref{coulomb})
are
%=================================================================
\be
\mbox{\boldmath{$ r$}}''- {r'\over r}\mbox{\boldmath{$ r$}}'
+ {\mu \over r} \mbox{\boldmath{$ r$}}
=
\mbox{\boldmath{$ 0$}},\quad
{1\over 2r^2}\mbox{\boldmath{$ r$}}'\cdot
\mbox{\boldmath{$ r$}}'- {\mu \over r} = h.
\label{coulomb'}
\ee
%=================================================================
Instead of using the variables $(x,y)$ it is convenient to make
the transformation \cite{LC}
%=================================================================
\be
\label{MAT1}
\left|\ba{c} x\\ y \ea\right| =\left| \ba{cc}u_1
& - u_2
\\
u_2 &u_1\ea \right| \left| \ba{c}{u_1}
\\
{u_2}\ea\right|
\quad \mbox{or }\quad
\mbox{\boldmath{$ r$}} =
L(\mbox{\boldmath{$ u$}}) \mbox{\boldmath{$ u$}}.
\ee
%=================================================================
>From the explicit form of these relations it follows that
$\mbox{\boldmath{$ r$}}'= 2L(\mbox{\boldmath{$ u$}})
\mbox{\boldmath{$ u$}}'$. The equations of motion are equivalent to
%=================================================================
\be
\mbox{\boldmath{$ u$}}''+ {{\mu \over 2} -
\mbox{\boldmath{$ u$}}'
\cdot\mbox{\boldmath{$ u$}}'\over \mbox{\boldmath{$ u$}}
\cdot \mbox{\boldmath{$ u$}}} \mbox{\boldmath{$ u$}}
= \mbox{\boldmath{$ u$}}, \quad
{\mu \over 2} = \mbox{\boldmath{$ u$}}'
\cdot\mbox{\boldmath{$ u$}}'-{h\over 2}
\mbox{\boldmath{$ u$}}\cdot  \mbox{\boldmath{$ u$}}.
\ee
%=================================================================
Consequently we have the regularized equation of motion
%=================================================================
$$
\mbox{\boldmath{$ u$}}''-{h\over 2}
\mbox{\boldmath{$ u$}}= \mbox{\boldmath{$ 0$}}.
$$
%=================================================================
This is essentially the equation for the harmonic oscillator if $h<0$.
The solution $u_1 = \alpha\cos(\omega s)$, $u_2 = \beta\sin(\omega s)$,
$\omega ^2=-h/2$ is equivalent to elliptical motion.

The relationship between the harmonic oscillator and the corresponding
Kepler problem can also be easily seen from the point of view of
Hamilton-Jacobi theory. Indeed the Hamiltonian can be written in the
two equivalent forms
%=================================================================
\be H={1\over 2}(p^2_x+p^2_y)+ {\mu \over \sqrt{x^2+y^2}}
= {1\over 8(u^2_1+u^2_2)}[p^2_{u_1}+p^2_{u_2}+8\mu ].
\label{hamiltonian}
\ee
%=================================================================
If we now write down the corresponding Hamilton-Jacobi equation
via the substitutions
%=================================================================
$$
p_{u_1}\rightarrow \partial _{u_1}S=S_{u_1}, \qquad
p_{u_2}\rightarrow \partial _{u_2}S=S_{u_2}
$$
%=================================================================
we obtain
%=================================================================
\be
S^2_{u_1}+S^2_{u_2}+8\mu -8E(u^2_1+u^2_2) = 0.
\label{HJequation}
\ee
%=================================================================
This is just the Hamilton-Jacobi equation for a mechanical system
with Hamiltonian
%=================================================================
$$
H'=p^2_{u_1}+p^2_{u_2}-8E(u^2_1+u^2_2)
$$
%=================================================================
and energy  $-8\mu $. (This is the pseudo-Coulomb problem, see
\cite{BKM,WT78}. Reference \cite{BKM} also obtains
(\ref{HJequation}) as an application of St\"ackel transform theory).

This transformation also achieves a regularization of the corresponding
quantum mechanical problem, which we call the {\it quantum Coulomb}
problem. Indeed, the Schr\"odinger equation in the
presence of the potential $V(r) = -\mu /r$ in two dimensions has the
form
%=================================================================
\be
-{1\over 2}(\partial ^2_x+\partial ^2_y)\Psi -
{\mu \over \sqrt{x^2+y^2}} \Psi  = E\Psi.
\label{Schrodinger}
\ee
%=================================================================
In the coordinates $(u_1,u_2)$, (\ref{Schrodinger}) becomes
\cite{KIB1}
%=================================================================
\be
(\partial ^2_{u_1}+\partial ^2_{u_2})\Phi +\{8\mu +
8E(u^2_1+u^2_2)\}\Phi = 0.
\label{Schrodinger1}
\ee
%=================================================================
Here, (\ref{Schrodinger1}) has all the appearances of the
Schr\"odinger equation in a oscillator potential $V(u_1,u_2) =
- 4E(u^2_1+u^2_2)$ and energy ${\cal E}= 4\mu$. Note that for
scattering state $E>0$ we have the repulsive oscillator potential
and for $E=0$ the free motion. For $E<0$ we get the attractive
oscillator potential and the corresponding bound state energy
spectrum can be easily computed from this reformulation of the
Coulomb problem, although the weight function for the inner product
is no longer the same \cite{KS,KIB1,BKM,WT81}. (Indeed, the Virial
Theorem states that for the Coulomb problem the change in weight
function does not alter the bound state spectrum, \cite{WT81}).
The wave functions have the form $\Phi =\varphi_1(u_1)\varphi_2(u_2)$
where the functions  $\varphi_\lambda $ satisfy
%=================================================================
$$
(\partial^2_{u_\lambda} + \kappa_\lambda + 8Eu^2_\lambda)\varphi_\lambda
=0, \quad
\lambda =1,2,\quad
\kappa_1+\kappa_2 = 8\mu.
$$
%=================================================================
The bound state eigenvalues are quantised according to
%=================================================================
\bea
\label{FLAT}
\kappa_\lambda = 2\sqrt{-2E}(2n_\lambda +1),
\quad
\lambda = 1,2,
\eea
%=================================================================
where $n_1,n_2$ are integers. Taking into account \cite{FLUG},
%=================================================================
$
\Phi(-u_1, -u_2) = (-1)^{n_1+n_2} \Phi(u_1, u_2)
$
%=================================================================
and using that $\Psi(x)$ is even in variable $u$:
%=================================================================
$
\Psi[x(u)] = \Psi[x(-u)],
$
%=================================================================
[because two points $(-u_1, -u_2)$ and $(u_1,u_2)$ in $u$-space
map to the same point in the plane $(x,y)$] we find
>from (\ref{FLAT}) the energy spectrum of the two dimensional Coulomb
system \cite{MAC,MPST}
%=================================================================
$$
E_N = -\frac{\mu^2}{2(N+\frac12)^2},
\qquad N=\frac{n_1+n_2}{2}=0,1,2\cdots.
$$
%=================================================================

It is well known that the regularizing transformations (\ref{MAT1})
that we have discussed for the Kepler and Coulomb problems in two
dimensional Euclidean spaces are also possible in the case of three
(Kustaanheimo-Stiefel transformation for mapping
R$_{4}$$\rightarrow$R$_{3}$) \cite{KS,IWAI1,TN,BP} and five
(Hurwitz transformation for mapping R$_{8}$$\rightarrow$R$_{5}$)
\cite{DMPST,KIB2,KIB3,HVT,IWAIS,MST1,PTOL} dimensions.
The only difference in these cases is that additional
constraints are required. These transformations have been employed
to solve many problems in classical and quantum mechanics
(see  \cite{KIB2} and references therein).

As in  flat space, the study of the Kepler-Coulomb system in  constant
curvature spaces has a long history. It was first
introduced in quantum mechanics by Schr\"odinger \cite{SCHR},
who used the factorization method to solve the Schr\"odinger equation
and to find the energy spectrum for the harmonic potential as an
analog of the Kepler-Coulomb potential on the three-dimensional
sphere. Later, two- and three-dimensional Coulomb and oscillator
systems were investigated by  many
authors in
\cite{HIG,LEEM,BOK,BIJ1,MPSV,DAS1,GRO1,KMJP,KMJP2,GRO2,PS}.

However, in spite of these achievements
the question of finding all  transformations that both  generalize the
Levi-Civita, Kustaanheimo-Steifel and Hurwitz transformations for
 spaces with constant curvature  and
preserve the Kepler-Coulomb and oscillator duality  has been open till
now. The answer to this question is a main aim of our paper.

The paper is organized as follows. In  \S2 we present the
transformations that generalize the flat space Levi-Civita
transformation and correspond to the map S$_{2C}$$\rightarrow$$S_{2}$
>from complex into real two-dimensional spheres. We show also that
this transformation establishes the correspondence between
Kepler-Coulomb and oscillator systems in classical and quantum
mechanics. In \S3,  in analogy with \S2, we construct
the Kustaanheimo-Steifel and Hurwitz transformation and show
Kepler-Coulomb and oscillator duality for  mappings
S$_{4C}$$\rightarrow$S$_{3}$ and S$_{8C}$$\rightarrow$S$_{5}$,
respectively.
 \S4 is devoted to a summary and discussion of our findings.
In the Appendix we give some formulas determining the connections between
Laplace-Beltrami operators and the volume elements in different spaces.

\section{The transformation on the 2-sphere}

The potential, which is the analogue of the Coulomb potential in  quantum mechanics and the gravitational potential
for the Kepler problem, is taken to be \cite{SCHR,HIG}
%=================================================================
\be
\label{POT1}
V = - \frac{\mu}{R}
{s_3\over \sqrt {s^2_1+s^2_2}},\label{coulombpot2}
\ee
%=================================================================
where ($s_1,s_2,s_3$) are the Cartesian coordinates in the ambient
Euclidean space and $R$ is the radius of the sphere
%=================================================================
$$
\mbox{\boldmath{$s$}} \cdot\mbox{\boldmath{$s$}}
\equiv  s^2_1+s^2_2+s^2_3=R^2.
$$
%=================================================================
(Note that $V = - \frac{\mu}{R}\cot\alpha$ where $\alpha$ is the arc
length distance from $\mbox{\boldmath{$s$}}$ to the north pole of
the sphere.  Furthermore, the leading term in the Laurent series
expansion in $\alpha$ about the north pole is $ - \frac{\mu}{R\alpha}$).

 This problem is easily transformed into a much
simpler one via the transform
%=================================================================
\bea
\label{TR1}
s_1 &=& i \sqrt{u^2_1+u^2_2+u^2_3}
\cdot\frac{u^2_1-u^2_2}{2u_3},
\nonumber\\
s_2 &=& i \sqrt{u^2_1+u^2_2+u^2_3}
\cdot\frac{u_1u_2}{u_3},
\\
s_3 &=&  \sqrt{u^2_1+u^2_2+u^2_3}
\cdot\left(u_3+\frac{u^2_1+u^2_2}{2u_3}\right),
\nonumber
\eea
%=================================================================
or in matrix form
%=================================================================
\bea
\left|\ba{c} s_1\\ s_2\\ s_3 \ea\right| =
\frac{\sqrt{u^2_1+u^2_2+u^2_3}}{2u_3}
\left| \ba{ccc}{iu_1}
& -{iu_2} & 0
\\
{iu_2} &{iu_1} & 0
\\
{u_1} &{u_2} & 2{u_3}
\ea \right| \left| \ba{c}{u_1}\\{u_2}
\\
{u_3}\ea\right|.
\eea
%=================================================================
The advantage of this transform is the Euler identity \cite{KIB1}
%=================================================================
\bea
s^2_1+s^2_2+s^2_3 = (u^2_1+u^2_2+u^2_3)^2,
\eea
%=================================================================
>from which we see that the point $\mbox{\boldmath{$u$}} =
(u_1,u_2,u_3)$ lies on the complex ``sphere" S$_{2C}$:
$u^2_1+u^2_2+u^2_3=D^2$ with the real radius $D$ if
$\mbox{ \boldmath{$s$}} = (s_1,s_2,s_3)$ lies on the real
sphere S$_2$ with radius $R$, and $R=D^2$.

In the general case the two-dimensional complex sphere S$_{2C}$
may be parametrized by four real variables
(the constraint $u^2_1+u^2_2+u^2_3=D^2$ includes two equations for
real and imaginary parts). The requirement of reality of the
Cartesian variables $s_i$ leads to  two more
equations and the formula (\ref{TR1}) corresponds to the mapping
>from a two-dimensional submanifold (or surface) in the complex sphere
S$_{2C}$ (four dimensional real space) to the sphere S$_2$.
To verify we introduce  ordinary spherical coordinates on S$_2$:
%=================================================================
\bea
s_1 = R \sin\chi \cos\varphi, \quad
s_2 = R \sin\chi \sin\varphi, \quad
s_3 = R \cos\chi.
\eea
%=================================================================
>From transformation (\ref{TR1}) we have
%=================================================================
\bea
\label{PAR}
\frac{s_3}{R} = \frac{1}{2}\left(\frac{u_3}{D} + \frac{D}{u_3}\right).
\eea
%=================================================================
Putting  $s_3=R\cos\chi$ in formula (\ref{PAR}) we get $u_3 = D
{e}^{i\chi}$ and then the corresponding points on the complex sphere
S$_{2C}$
are
%=================================================================
\bea
\label{CU}
u_1 = D \sqrt{1-{e}^{2i\chi}} \cos{\frac{\varphi}{2}}, \quad
u_2 = D \sqrt{1-{e}^{2i\chi}} \sin{\frac{\varphi}{2}}, \quad
u_3 = D {e}^{i\chi}
\eea
%=================================================================
where $0\leq\chi\leq\pi$, \, $0\leq\varphi\leq 4\pi$. Note  that
the transformation  (\ref{TR1}) is not one to one; two points
$(-u_1, -u_2, u_3)$ and $(u_1, u_2, u_3)$ on the sphere in $u$-space
correspond to  one point on the sphere in $s$-space. Thus, when the
variables $(u_1,u_2,u_3)$ cover the sphere in $u$-space, the variables
$s_i$ cover the sphere in $s$-space twice.

Let us now introduce  nonhomogeneous coordinates according to
\cite{IPSW}
%=================================================================
\bea
{\bar s}_i = R \frac{s_i}{s_3}, \qquad
{\bar u}_i = D \frac{u_i}{u_3}, \qquad D^2 = R,
\qquad i=1,2.
\eea
%=================================================================
Then formula (\ref{TR1}) transforms to
%=================================================================
\bea
\label{TR2}
{\bar s}_1 =  \frac{i({\bar u}^2_1-{\bar u}^2_2)}
{2(1+\frac{{\bar u}^2_1+{\bar u}^2_2}{2D^2})},
\qquad
{\bar s}_2 =  \frac{i{\bar u}_1 \cdot {\bar u}_2}
{(1+\frac{{\bar u}^2_1+{\bar u}^2_2}{2D^2})}.
\eea
%=================================================================
In the contraction limit $D\rightarrow\infty$ we obtain
%=================================================================
\bea
\label{TR3}
{\bar s}_1 =  i \frac{{\bar u}^2_1-{\bar u}^2_2}{2},
\qquad
{\bar s}_2 =  i {\bar u}_1 \cdot {\bar u}_2,
\eea
%=================================================================
which coincides with the flat space Levi-Civita transformation
(\ref{MAT1}) up to the additional mapping
${\bar u}_i\rightarrow e^{-i\frac{\pi}{4}}\sqrt{2}{\tilde u_i}$.

The relationship between the infinitesimal distances is
%=================================================================
\be
d\mbox{\boldmath{$s$}} \cdot d\mbox{\boldmath{$s$}} =
(u^2_1+u^2_2+u^2_3)\left[\frac{(\mbox{\boldmath{$u$}}
\cdot d\mbox{\boldmath{$u$}})^2}{u^2_3} -
\left({u^2_1+u^2_2\over u^2_3}\right)
d \mbox{\boldmath{$u$}} \cdot d\mbox{\boldmath{$u$}}
\right]
+ 3 (\mbox{\boldmath{$u$}}\cdot d\mbox{\boldmath{$u$}})^2.
\ee
%=================================================================
Thus, when restricted to the sphere, the infinitesimal distances are
related by
%=================================================================
\be
\frac{d\mbox{\boldmath{$s$}} \cdot d\mbox{\boldmath{$s$}}}{R}
= - \left({u^2_1+u^2_2\over u^2_3}\right)
d\mbox{\boldmath{$u$}} \cdot d\mbox{\boldmath{$u$}},
\ee
%=================================================================
and we see that as in  flat space the transformation (\ref{TR1})
is conformal.

\subsection{Classical motion}

Just as in the case of Euclidean space, the  classical
equations of motion under the influence of  a Coulomb potential can be
simplified. The classical equations are
%=================================================================
\be
\ddot{\mbox{\boldmath{$ s$}}} =
- (\dot{\mbox{\boldmath{$s$}}}
\cdot\dot{\mbox{\boldmath{$ s$}}})\mbox{\boldmath{$s$}}
- \mbox{\boldmath{$\nabla $}}V, \label{newton1}
\ee
%=================================================================
where the first term on the right hand side is the centripetal force
term, corresponding to the constraint of the motion to the sphere,
and the potential satisfies
%=================================================================
\be
\mbox{\boldmath{$s$}}\cdot \mbox{\boldmath{$ \nabla $}}V = 0.
\label{newton2}
\ee
%=================================================================
Here,
$\dot{\mbox{\boldmath{$s$}}}=\frac{d}{dt}\mbox{\boldmath{$s$}}$.
(In studying (\ref{newton1}) and (\ref{newton2}) we initially regard
the coordinates $\mbox{\boldmath{$s$}}$ as unconstrained and then
restrict our attention to solutions on the sphere).
In the case of  potential (\ref{coulombpot2}) these
equations become
%=================================================================
$$
{d^2\over dt^2}s_j = - s_j(\dot{\mbox{\boldmath{$s$}}}
\cdot\dot{\mbox{\boldmath{$s$}}}) -
\frac{\mu}{R}{s_js_3\over (s^2_1+s^2_2)^\frac32} ,
\quad  j=1,2
$$
%=================================================================
$$
{d^2\over dt^2}s_3 = - s_3 (\dot{\mbox{\boldmath{$s$}}}
\cdot\dot{\mbox{\boldmath{$s$}}}) + \frac{\mu}{R}
{1 \over (s^2_1+s^2_2)^\frac12},
$$
%=================================================================
subject to the constraints
%=================================================================
\be
\mbox{\boldmath{$s$}} \cdot\mbox{\boldmath{$s$}} = R^2
\ee
%=================================================================
and its differential consequences
%=================================================================
$$
\mbox{\boldmath{$s$}} \cdot\dot{\mbox{\boldmath{$s$}}} = 0,
\quad
\mbox{\boldmath{$s$}} \cdot\ddot{\mbox{\boldmath{$s$}}}
+ \dot{\mbox{\boldmath{$s$}}}
\cdot\dot{\mbox{\boldmath{$s$}}} = 0.
$$
%=================================================================
>From the equations of motion we immediately deduce the energy
integral
%=================================================================
\be
{1\over 2}\dot{\mbox{\boldmath{$s$}}}
\cdot\dot{\mbox{\boldmath{$s$}}}+V = E.
\label{energy}
\ee
%=================================================================
We choose a new variable $\tau $ such that
%=================================================================
$$
{d\tau \over dt}= \frac{1}{D^2} \cdot {u^2_3\over u^2_1+u^2_2}.
$$
%=================================================================
In terms of the variables $\tau$ and $u_i$, the equations of
motion can now be written in the form
%=================================================================
\bea
(u'_1)^2+(u'_2)^2+(u'_3)^2
- 2D^2 (E+\frac{i\mu}{D^2}) +
\frac{2D^4}{u^2_3}(E-\frac{i\mu}{D^2})& =&0\\
{u''_1}+2(E+\frac{i\mu}{D^2}) u_1 = 0, \quad
{u''_2}+2(E+\frac{i\mu}{D^2}) u_2 &=&0
\\
{u''_3}+2(E+\frac{i\mu}{D^2}) u_3 -
\frac{2D^4}{u^3_3}(E-\frac{i\mu}{D^2})&=&0,
\eea
%=================================================================
subject to the constraint  $\mbox{\boldmath{$u$}} \cdot
\mbox{\boldmath{$u$}} = D^2$ and its differential consequences
$\mbox{\boldmath{$u$}} \cdot \mbox{\boldmath{$u$}}'= 0$,
$\mbox{\boldmath{$u$}} \cdot \mbox{\boldmath{$u$}}''+
\mbox{\boldmath{$u$}}' \cdot \mbox{\boldmath{$u$}}'= 0$,
where $u'_i={du_i\over d\tau }$. These equations are equivalent
to the equations of motion we would obtain by choosing the
Hamiltonian
%=================================================================
\be
H = {1\over 2}(p^2_{u_1}+p^2_{u_2}+p^2_{u_3})
- (E+\frac{i\mu}{D^2})(u^2_1+u^2_2+u^2_3) + \frac{D^4}{u^2_3}
(E-\frac{i\mu}{D^2}),
\ee
%=================================================================
regarding the variables $u_i$ as independent and using the variable
$\tau$ as time. In fact, to solve the classical mechanical
problem from the point of view of the Hamilton-Jacobi equation,
we use the relation
%=================================================================
\bea
\frac12(p^2_{s_1} + p^2_{s_2} + p^2_{s_3})
- \frac{\mu}{R} \frac{s_3}{\sqrt{s^2_1 + s^2_2}} - E \equiv
\nonumber\\[2mm]
- {u^2_3\over u^2_1+u^2_2}
\left[\frac{1}{2D^2}(p^2_{u_1} + p^2_{u_2} + p^2_{u_3})
- (i\frac{\mu}{D^2} + E) + \frac{D^2}{u^2_3}
(E-\frac{i\mu}{D^2})\right]
= 0,
\eea
%=================================================================
together with the substitutions $p_{u_i}={\partial S\over \partial u_i}$
and $p_{s_j}={\partial S\over \partial s_j}$ to obtain the
Hamilton-Jacobi equations
%=================================================================
\be
\left({\partial S\over\partial s_1}\right)^2+
\left({\partial S\over \partial s_2}\right)^2
+\left({\partial S\over \partial s_3}\right)^2
- \frac{2\mu}{R}\frac{s_3}{\sqrt{s^2_1+s^2_2}}- 2E = 0,
\ee
%=================================================================
\be
\left({\partial S\over \partial u_1}\right)^2
+\left({\partial S\over \partial u_2}\right)^2
+\left({\partial S\over \partial u_3}\right)^2
- 2D^2(\frac{i\mu}{D^2} + E) + \frac{2D^4}{u^2_3}
(E-\frac{i\mu}{D^2}) = 0.
\ee
%=================================================================
This last equation can be solved by separation of variables in the
spherical coordinates on the complex sphere S$_{2C}$
(\ref{CU}).

\subsection{Quantum motion}

 If we write the Schr\"odinger
equation on the sphere for the Coulomb potential (\ref{POT1})
%=================================================================
\bea
\label{KC1}
\frac{1}{2}\Delta_s \Psi +
\left(E + \frac{\mu}{R}{s_3\over\sqrt{s^2_1+s^2_2}}
\right)\Psi = 0,
\eea
%=================================================================
and use the transformation (12), we obtain
[see formula (\ref{LB22})]
%=================================================================
\bea
\label{OS1}
\frac{1}{2}\Delta_u \psi + \left({\cal E} -
\frac{\omega^2 D^2}{2} \frac{u_1^2+u_2^2}{u_3^2}\right)\psi = 0
\eea
%=================================================================
where
%=================================================================
\bea
\label{EOS1}
{\cal E} = 2i\mu,  \qquad   \omega^2 = 2(E - \frac{i\mu}{D^2}).
\eea
%=================================================================
Thus we see that the Coulomb problem on the real sphere {S}$_{2}$
is equivalent to the corresponding quantum mechanical problem on
the complex sphere S$_{2C}$ with the oscillator potential
(Higgs oscillator \cite{HIG,DAS1,GRO1}) and energy
$2i\mu$, but with an altered inner product (see the Appendix).

Let us consider the Schr\"odinger equation (\ref{OS1}). Using the
complex spherical coordinates (\ref{CU}) we obtain
%=================================================================
\bea
\label{TDE}
\frac{1}{\sin\chi}\frac{\partial}{\partial\chi}
\sin\chi \frac{\partial \psi}{\partial\chi}
+ \frac{1}{\sin^2\chi}\frac{\partial\psi}{\partial\varphi}
+  \left\{\omega^2D^4 - i{\cal E}D^2 \frac{{\mbox e}^{i\chi}}
{\sin\chi}\right\} \psi =0.
\eea
%=================================================================
To solve this equation we first complexify the Coulomb coupling constant
$\mu$ by setting $k = i\mu$ in the formulas for ${\cal E}$ and $\omega$
%=================================================================
\bea
\label{MU1}
{\cal E}  = 2k, \qquad
\omega^2 = 2(E - \frac{k}{D^2}).
\eea
%=================================================================
Further, we analytically continue  the variable $\chi$
into the complex domain {\bf G}: $0\leq {\mbox Re} \chi \leq \pi$
and $0\leq {\mbox Im} \chi < \infty$, (see Fig.1) and pass from
the variable $\chi$ to  $\vartheta$, defined by
%=================================================================
\bea
\label{COM}
{\mbox e}^{i\chi} = \cos\vartheta.
\eea
%=================================================================
\unitlength=1.00mm
\special{em:linewidth 0.4pt}
\linethickness{0.4pt}
\begin{picture}(102.00,126.60)
\put(14.00,44.00){\makebox(0,0)[cc]{0}}
\put(12.00,120.00){\makebox(0,0)[cc]{Im $\chi$}}
\put(102.00,42.00){\makebox(0,0)[cc]{Re $\chi$}}
\put(98.00,50.00){\line(-1,0){91.00}}
\put(65.00,50.00){\line(0,1){67.00}}
\put(65.00,43.00){\makebox(0,0)[cc]{$\pi$}}
\put(65.00,119.00){\line(-1,0){4.00}}
\put(59.00,119.00){\line(-1,0){4.00}}
\put(52.00,119.00){\line(-1,0){5.00}}
\put(44.00,119.00){\line(-1,0){5.00}}
\put(36.00,119.00){\line(-1,0){5.00}}
\put(28.00,119.00){\line(-1,0){5.00}}
\put(41.00,86.00){\makebox(0,0)[cc]{{\bf G}}}
\put(74.00,86.00){\makebox(0,0)[cc]{G}}
\put(96.00,50.00){\vector(1,0){2.00}}
\put(39.00,50.00){\vector(1,0){3.00}}
\put(65.00,84.00){\vector(0,1){1.00}}
\put(44.00,119.00){\vector(-1,0){2.00}}
\put(20.00,86.00){\vector(0,-1){2.00}}
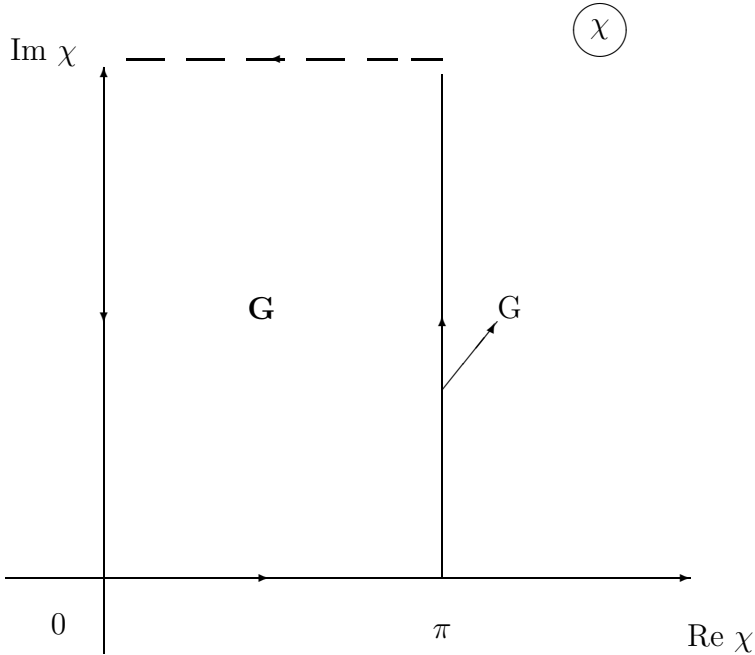
\put(-5.00,22.00){\makebox(0,0)[lc]{{\bf Figure 1:}\ \  Domain {\bf G} =\{ $0\leq$ Re $\chi\leq \pi$; $0\leq$ Im $\chi< \infty$\} on the complex plane of $\chi$.}}
\put(20.00,40.00){\line(0,1){78.00}}
\put(86.00,123.00){\circle{7.21}}
\put(86.00,123.00){\makebox(0,0)[cc]{$\chi$}}
\put(65.00,75.00){\line(4,5){7.33}}
\put(71.00,82.00){\vector(1,2){1.00}}
\put(20.00,118.00){\vector(0,1){0.00}}
\end{picture}
%=================================================================

For real  $\theta$ this substitution is possible if {Re}$\chi=0$ or Re$\chi=\pi$ and Im$\chi\in (0, \infty)$,
which corresponds to the motion on the upper ($0\leq \vartheta \leq
\frac{\pi}{2}$) or lower ($\frac{\pi}{2}\leq \vartheta \leq \pi$)
hemispheres of the real sphere. In any case  conditions (\ref{COM})
and (\ref{MU1}) translate the oscillator problem from the complex to
the real sphere with  spherical coordinates ($\vartheta, \varphi/2$).
In these coordinates we can rewrite  (\ref{TDE}) in the form
%=================================================================
\bea
\label{TDE1}
\frac{1}{\sin\vartheta}\frac{\partial}{\partial\vartheta}
\sin\vartheta \frac{\partial \psi}{\partial\vartheta}
+
\frac{4}{\sin^2\vartheta}\frac{\partial^2\psi}{\partial\varphi^2}
+
\left\{(2{\cal E} D^2 + \omega^2 D^4) - \frac{\omega^2D^4}
{\cos^2\vartheta}\right\} \psi = 0.
\eea
%=================================================================
Using the separation of variables ansatz
%=================================================================
\bea
\label{WAVE1}
\psi(\vartheta, \varphi) = R(\vartheta) \,
\frac{{\mbox e}^{im\frac{\varphi}{2}}}{\sqrt{2\pi}},
\qquad m = 0, \pm 1, \pm 2, ...
\eea
%=================================================================
we obtain
%=================================================================
\bea
\label{QUASI1}
\frac{1}{\sin\vartheta}\frac{d}{d\vartheta}
\sin\vartheta \frac{d R}{d\vartheta}
+
\left\{(2{\cal E} D^2 + \omega^2 D^4) - \frac{\omega^2D^4}
{\cos^2\vartheta} - \frac{m^2}{\sin^2\vartheta}\right\} R = 0.
\eea
%=================================================================
The corresponding solution regular at the points $\vartheta = 0,
\pi/2$ takes the form \cite{FLUG}
%=================================================================
\bea
\label{RF22}
R_{n_r m}(\vartheta)
&=&
C_{n_r m}(\nu) \, (\sin\vartheta)^{|m|}\,
(\cos\vartheta)^{\nu+\frac12} \,
{_2F_1}(-n_r, n_r+\nu+|m|+1; \, |m|+1;
\sin^2\vartheta)
\nonumber\\[2mm]
&=&
C_{n_r m}(\nu) \,
\frac{(n_r)!|m|!}{(n_r+|m|)!}\,
(\sin\vartheta)^{|m|}\, (\cos\vartheta)^{\nu+\frac12}\,
P^{(|m|, \nu)}_{n_r} (\cos 2\vartheta)
\eea
%=================================================================
with energy spectrum given by expression
%=================================================================
\bea
\label{EN1}
{\cal E}  = \frac{1}{2D^2}[(n+1)(n+2) + (2\nu-1)(n+1)],
\,\,\,\,\,\,\,\,
\nu = \left(\omega^2D^4+\frac14\right)^{\frac12}
\eea
%=================================================================
where $C_{n_r m}(\nu)$ is the normalization constant,
$P^{(\alpha, \beta)}_{n}(x)$ is a Jacobi polynomial, $n_r=0,1,2...$
is the ``radial" and  $n=2n_r+|m|$ is the principal quantum number.

To compute the normalization constant $C_{n_r m}(\nu)$ for the
reduced system we require that the wave function (\ref{WAVE1}) satisfy
the normalization condition (see the Appendix):
%================================================================
\bea
\label{NOR1}
- \frac{D^2}{2} \,
\int_{S_{2C}}\, \psi_{n_r m}\psi^{\diamond}_{n_r m}
\frac{u_1^2+u_2^2}{u_3^2}\, d v(u) =
D^4 \, \int_{0}^{\pi} R_{n_r m} R^{\diamond}_{n_r m} \sin\chi d\chi
= 1
\eea
%================================================================
where the symbol "${\diamond}$" means the complex conjugate together
with the inversion $\chi\rightarrow -\chi$, i.e.
$\psi^{\diamond}(\chi,\varphi) = \psi^{*}(-\chi, \varphi)$.
[We choose the scalar product as $\psi^{\diamond}\psi$ because
for real $\omega^2$ and ${\cal E}$ the function
$\psi^{\diamond}(\chi, \vartheta)$ also belongs to the solution space of
  (\ref{TDE}).]

Consider now the integral over  contour $G$ in the complex
plane of variable $\chi$ (see Fig.1)
%================================================================
\bea
\label{NOR2}
\oint R_{n_rm} R^{\diamond}_{n_r m} \sin\chi d\chi
&=&
\int_{0}^{\pi} R_{n_rm} R^{\diamond}_{n_r m} \sin\chi d\chi
+
\int_{\pi}^{\pi+i\infty} R_{n_rm} R^{\diamond}_{n_r m} \sin\chi d\chi
\nonumber\\[2mm]
&+&
\int_{\pi+i\infty}^{i\infty} R_{n_rm} R^{\diamond}_{n_r m} \sin\chi d\chi
+
\int_{i\infty}^0 R_{n_rm} R^{\diamond}_{n_r m} \sin\chi d\chi.
\eea
%================================================================
Using the facts that
the integrand  vanishes as $e^{2i\nu\chi}$ and
 that  $R_{n_rm}(\chi)$ is regular in the domain
{\bf G} (see Fig.1), then according to the Cauchy theorem we have
%================================================================
\bea
\label{NOR3}
\int_{0}^{\pi} R_{n_rm} R^{\diamond}_{n_r m} \sin\chi d\chi
&=&
\int_0^{i\infty} R_{n_rm} R^{\diamond}_{n_r m} \sin\chi d\chi
-
\int_{\pi}^{\pi+i\infty} R_{n_rm} R^{\diamond}_{n_r m} \sin\chi d\chi
\nonumber\\[3mm]
&=&
\left[1-e^{2i\pi(\nu+\frac12)}\right]\,
\int_0^{i\infty} R_{n_rm} R^{\diamond}_{n_r m} \sin\chi d\chi.
\eea
%================================================================
Making the substitution (\ref{CU}) in the right integral
of eq. (\ref{NOR3}), we find
%================================================================
\bea
\int_{0}^{\pi} R_{n_rm} R^{\diamond}_{n_r m} \sin\chi d\chi
=
\left[1-e^{2i\pi(\nu+\frac12)}\right]\,
\int_{0}^{\frac{\pi}{2}} [R_{n_rm}]^2 \sin\vartheta \tan^2\vartheta
d\vartheta.
\eea
%================================================================
Using the following formulas for integration of the two
Jacobi polynomials \cite{BER}
%================================================================
\bea
\int_{-1}^{1}\, (1-x)^{\alpha}(1+x)^{\beta}
[P^{(\alpha, \beta)}_{n}(x)]^2 dx
&=&
\frac{2^{\alpha+\beta+1}\Gamma(n+\alpha+1)\Gamma(n+\beta+1)}
{(2n+\alpha+\beta+1)n!\Gamma(n+\alpha+\beta+1)}
\nonumber\\[2mm]
\int_{-1}^{1}\, (1-x)^{\alpha}(1+x)^{\beta-1}
[P^{(\alpha, \beta)}_{n}(x)]^2 dx
&=&
\frac{2^{\alpha+\beta}\Gamma(n+\alpha+1)\Gamma(n+\beta+1)}
{(\beta) n!\Gamma(n+\alpha+\beta+1)}
\nonumber
\eea
%================================================================
we find
%===============================================================
\bea
\label{SSS1}
C_{n_r m}(\nu)
=
\frac{2}
{(|m|)!}
\,\,
\sqrt{\frac{-\nu(\nu+2n_r+|m|+1)\, (n_r+|m|)! \Gamma(|m|+n_r+\nu+1)}
{D^4 [1-e^{2i\pi(\nu+\frac12)}](2n_r+|m|+1)\,(n_r)!
\Gamma(n_r+\nu+1)}}.
\eea
%================================================================
The wave function
$\psi(\vartheta, \varphi) \equiv \psi_{n_r m}(\vartheta, \varphi)$
is then given by eqs. (\ref{WAVE1}), (\ref{RF22}) and (\ref{SSS1}).

Now we can construct the Coulomb wave functions and eigenvalue
spectrum. From transformation
%=================================================================
\bea
\psi_{n_r m}(\vartheta, \varphi+2\pi) =  e^{im\pi}
\psi_{n_r m}(\vartheta, \varphi)
\eea
%=================================================================
and requirement of $2\pi$ periodicity for the wave functions
(\ref{WAVE1}) we see  that only even azimuthal angular
momentum states of the oscillator correspond to the reduced system.
Then, introducing new angular and principal quantum numbers $M$
and $N$ by the condition
%=================================================================
\bea
n=2n_r+|m|=2n_r+2|M|=2N,
\qquad  N=0,1,2,... ,
\,\,\,\,\,\,\,
|M|=0,1,2,.. N,
\eea
%=================================================================
comparing (\ref{MU1}) with  expression (\ref{EN1}) for the oscillator
energy spectrum, and putting  $k=i\mu$, we find the energy
spectrum for reduced systems
%=================================================================
\bea
E_{N} = \frac{N(N+1)}{2R^2} -
\frac{\mu^2}{2(N+\frac12)^2}.
\eea
%=================================================================
This formula coincides with that obtained from other methods
in works \cite{HIG,DAS1,GRO1}.

Transforming $\vartheta$ back to the variable $\chi$ by (\ref{COM}),
we see that (\ref{EN1}) and (\ref{MU1}) imply
%=================================================================
$$
\nu = i\sigma - (N+ \frac12),
\qquad
\sigma = \frac{\mu R}{N+\frac12}.
$$
%=================================================================
Using
%=================================================================
\bea
\frac{\Gamma(1/2+|M|+i\sigma)}{\Gamma(1/2-|M|+i\sigma)}
&=& (-1)^{|M|} \frac{|\Gamma(1/2+|M|+i\sigma)|^2}
{|\Gamma(1/2+i\sigma)|^2}
\nonumber\\[2mm]
&=&
\frac{(-1)^{|M|}}{\pi}\,\cosh\sigma\pi
\, |\Gamma(1/2+|M|+i\sigma)|^2
\eea
%=================================================================
we easily get from (\ref{WAVE1}), (\ref{RF22}) and (\ref{SSS1}) the
eigenfunction of Schr\"odinger equation (\ref{KC1})
%=================================================================
\bea
\label{SS22}
\Psi_{NM}(\chi, \varphi)
&=&
C_{NM}(\sigma) \,
e^{-i\chi (N-|M|-i\sigma)}
\,
(\sin\chi)^{|M|}\,
\,
\nonumber\\[2mm]
&\times&  _2F_1(-N+|M|, \, |M|+i\sigma+\frac12; \,
2|M|+1; \, 1-e^{2i\chi})\,
\, \frac{e^{iM\varphi}}{\sqrt{2\pi}}.
\eea
%=================================================================
where now
%=================================================================
\bea
C_{NM}(\sigma) =
\frac{2^{|M|}}{R (2|M| )!}
\sqrt{\frac{[(N+\frac12)^2+\sigma^2](N+|M|)!}
{\pi(N +\frac12)(N-|M|)!}}\, e^{\frac{\sigma\pi}{2}}
\,
|\Gamma(|M|+1/2+i\sigma)|
\eea
%=================================================================
By  direct calculation it may be shown that the Coulomb wave function
(\ref{SS22}) satisfies the normalization condition
%================================================================
$$
R^2 \, \int_{0}^{\pi}\sin\chi d\chi \int_{0}^{2\pi} d\varphi
\, \Psi_{N M} \Psi^{*}_{N' M'}
= \delta_{NN'} \delta_{MM'}.
$$
%================================================================
Thus, by reduction from the two-dimensional quantum oscillator on the
complex sphere we have constructed the wave function and energy
spectrum for the Coulomb problem on the two-dimensional real
sphere $S_2$.  Formula (\ref{SS22}) for Coulomb
wave functions on the two-dimensional sphere is new.

Now let us consider the flat space contraction.
In the contraction limit $R\rightarrow\infty$ the energy spectrum
for finite  $N$ goes to the discrete energy
spectrum of the two-dimensional hydrogen atom  \cite{MAC,MPST}
%=================================================================
$$
\lim_{R\rightarrow\infty} E_N(R) = -\frac{\mu^2}{2(N+\frac12)^2},
\qquad N=0,1,...
$$
%=================================================================
In the limit $R\rightarrow\infty$, putting
$\tan\chi \sim \chi \sim \frac{r}{R}$, where $r$ is the radius-vector
in the two-dimensional tangent plane and using the asymptotic
formulas \cite{BE}
%=================================================================
$$
\lim_{\scriptstyle R\rightarrow\infty\atop\scriptstyle
\chi\rightarrow 0}\,
{_2F_1}(-N+|M|, \, |M|+i\sigma+\frac12; \,
2|M|+1; \, 1-e^{2i\chi})
$$
\bea
\label{CONTR1}
&=&
{_1F_1}(-N+|M|,\, 2|M|+1; \, \frac{2\mu r}{N+\frac12})
\nonumber\\[2mm]
\lim_{|y|\rightarrow\infty} \,
|\Gamma(x+iy)| \, e^{\frac{\pi}{2} y}\,
|y|^{\frac12-x}
&=&
\sqrt{2\pi},
\qquad
\lim_{z\rightarrow\infty} \,
\frac{\Gamma(x+\alpha)}{\Gamma(x+\beta)}
= z^{\alpha-\beta},
\eea
%=================================================================
we obtain the well known Coulomb wave function with correct
normalization factor \cite{MPST}
%=================================================================
\bea
\lim_{\scriptstyle R\rightarrow\infty\atop\scriptstyle
\chi\rightarrow 0}
\Psi_{NM}(\chi, \varphi)
&=&
\frac{\mu \sqrt{2}}{(N+\frac12)^{3/2}}\,
\sqrt{\frac{(N+|M|)!}{N-|M|)!}}\,
\left(\frac{2\mu r}{N+\frac12}\right)^{|M|}\,
\frac{e^{-\frac{\mu r}{N+\frac12}}}{(2|M|)!}\,
\nonumber\\[2mm]
&\times&
_1F_1(-N+|M|,\, 2|M|+1; \, \frac{2\mu r}{N+\frac12})
\,
\frac{e^{iM\varphi}}{\sqrt{2\pi}}.
\eea
%=================================================================
In the case for large $R$ and $N$ such that $N\sim kR$,
(where $k$ is constant) we obtain the formula for continuous spectrum:
$E = k^2/2$. Now taking into account that $\sigma \sim \frac{\mu}{k}$
and using the asymptotic relation  (\ref{CONTR1}),
we have
%=================================================================
\bea
\lim_{\scriptstyle R\rightarrow\infty\atop\scriptstyle
\chi\rightarrow 0}
\sqrt{R} \,
\Psi_{NM}(\chi, \varphi)
&=&
\sqrt{\frac{k}{\pi}}\,
e^{\frac{\pi \mu}{2k}}\,
|\Gamma(|M|+1/2+i\mu/k)|\,
\frac{(2kr)^{|M|}}{(2|M|)!}\,
e^{-ikr}
\nonumber\\[2mm]
&\times&
_1F_1(|M|+\frac{i\mu}{k}+\frac12; \,
2|M|+1; \, {2ikr})
\,
\frac{e^{iM\varphi}}{\sqrt{2\pi}},
\eea
%=================================================================
which  coinsides with the formula for the two-dimensional Coulomb scattering
wave function in polar coordinates \cite{DPST}.

\section{The three and five dimensional Kepler - Coulomb problems}

In complete analogy with the three- and five-dimensional Euclidean
case the corresponding regularizing transformations exist for the
Kepler and Coulomb problems in spheres of dimension 3 and 5.
Indeed if we consider motion on the sphere of dimension $n$
then the classical equations of motion in the presence of a
potential are just (\ref{newton1}), (\ref{newton2}) again, where now
%=================================================================
\be
\mbox{\boldmath{$s$}} = (s_1,\cdots,s_{n+1}),
\ee
%=================================================================
subject to the constraints
%=================================================================
\be
\mbox{\boldmath{$s$}} \cdot\mbox{\boldmath{$s$}} = R^2
\ee
%=================================================================
and its differential consequences
%=================================================================
$$
\mbox{\boldmath{$s$}} \cdot\dot{\mbox{\boldmath{$s$}}}=0,\quad
\mbox{\boldmath{$s$}}
\cdot\ddot{\mbox{\boldmath{$s$}}} +
\dot{\mbox{\boldmath{$s$}}}
\cdot\mbox{\boldmath{$\dot s$}} = 0.
$$
%=================================================================

If we choose our potential to be
%=================================================================
\be
\label{POT2}
V= - \frac{\mu}{R}  {s_{n+1}\over \sqrt {s^2_1+..+s^2_n}},
\ee
%=================================================================
these equations assume the form
%=================================================================
\be
{d^2\over dt^2}s_j= - s_j\dot{\mbox{\boldmath{$s$}}} \cdot
\dot{\mbox{\boldmath{$s$}}} - \frac{\mu}{R}
{s_js_{n+1}\over(\mbox{\boldmath{$s$}}
\cdot\mbox{\boldmath{$s$}})^\frac32} , \quad j=1, \dots, n,
\ee
%=================================================================
\be
{d^2\over dt^2}s_{n+1} = -
s_{n+1}\dot{\mbox{\boldmath{$s$}}}
\cdot\dot{\mbox{\boldmath{$s$}}}+
{\mu\over R(\mbox{\boldmath{$s$}}
\cdot\mbox{\boldmath{$s$}})^\frac12 }.
\nonumber
\ee
%=================================================================
The energy integral again has the form  (\ref{energy}).

We are particularly interested in  dimensions $n=3,5$.
We deal with each of these cases separately.

\subsection{Generalized KS transformation}

For $n=3$ we choose the $u_j$ coordinates in five dimensional space
according to
%=================================================================
\bea
\label{KS}
s_1&=&i\sqrt{u^2_1+u^2_2+u^2_3+u^2_4+u^2_5} \cdot
\frac{u_1 u_3 + u_2 u_4}{u_5},
\nonumber
\\[2mm]
s_2&=&i\sqrt{u^2_1+u^2_2+u^2_3+u^2_4+u^2_5} \cdot
\frac{u_2 u_3 - u_1u_4}{u_5},
\nonumber
\\[2mm]
s_3&=&i\sqrt{u^2_1+u^2_2+u^2_3+u^2_4+u^2_5} \cdot
\frac{u^2_1+u^2_2-u^2_3-u^2_4}{2u_5},
\nonumber
\\[2mm]
s_4&=& \sqrt{u^2_1+u^2_2+u^2_3+u^2_4+u^2_5} \cdot
\left(u_5+\frac{u^2_1+u^2_2+u^2_3+u^2_4}{ 2u_5}\right).
\eea
%=================================================================
The basic identity is
%=================================================================
$$
s^2_1+s^2_2+s^2_3+s^2_4 = (u^2_1+u^2_2+u^2_3+u^2_4+u^2_5)^2,
$$
%=================================================================
and the basic relationship for the infinitesimal distances
is
%=================================================================
\be
ds^2_1+ds^2_2+ds^2_3+ds^2_4 =
\ee
%=================================================================
$$
-{D^2\over u^2_5}
\{(u^2_1+u^2_2+u^2_3+u^2_4)[du^2_1+du^2_2+du^2_3+du^2_4+du^2_5]
$$
%=================================================================
$$
+ (u_4du_3 - u_3du_4 + u_2du_1 - u_1du_2)^2\},
$$
%=================================================================
where the constraint for mapping between the 3-sphere:
$\sum_{i=1}^{4} s_i^2=R^2$ and the complex 4-sphere:
$\sum_{i=1}^{5} u_i^2=D^2$  is clearly
%=================================================================
\be
u_4du_3 - u_3du_4 + u_2du_1 - u_1du_2 = 0.
\ee
%=================================================================
In this section we will use the Eulerian spherical coordinates on
the complex 4-sphere S$_{4C}$
%=================================================================
\bea
\label{COOR34}
\begin{array}{ll}
u_1 = D \sqrt{1-{\mbox e}^{2i\chi}}\cos{\frac{\beta}{2}}
\cos{\frac{\alpha+\gamma}{2}}, \,\,\,&
u_2 = D \sqrt{1-{\mbox e}^{2i\chi}}\cos{\frac{\beta}{2}}
\sin{\frac{\alpha+\gamma}{2}},
\\[3mm]
u_3 =  D \sqrt{1-{\mbox e}^{2i\chi}}\sin{\frac{\beta}{2}}
\cos{\frac{\alpha-\gamma}{2}},  \,\,\,&
u_4 = D \sqrt{1-{\mbox e}^{2i\chi}}\sin{\frac{\beta}{2}}
\sin{\frac{\alpha-\gamma}{2}},
\\[3mm]
u_5=D{\mbox e}^{i\chi},
\end{array}
\eea
%=================================================================
where the ranges of the variables are given by
%=================================================================
$$
0 \leq \chi \leq \pi, \qquad
0 \leq \beta \leq \pi, \qquad
0 \leq \alpha < 2 \pi, \qquad
0 \leq \gamma < 4\pi.
$$
%=================================================================
The corresponding  spherical coordinates
on $S_3$ are
%=================================================================
\bea
\begin{array}{ll}
s_1=R\sin\chi\sin\beta\cos\alpha, &
s_2 = R\sin\chi\sin\beta\sin\alpha,
\\[2mm]
s_3=R\sin\chi\cos\beta, &
s_4 = R\cos\chi.
\nonumber
\end{array}
\eea
%=================================================================

\subsubsection{Classical motion}

In analogy with our previous analysis we choose a new variable
$\tau $ according to
%=================================================================
$$
{d\tau \over dt} = \frac{1}{D^2} \,
{u^2_5\over u^2_1+u^2_2+u^2_3+u^2_4}.
$$
%=================================================================
In the $u$ coordinates the equations of motion can be written as
%=================================================================
$$
(u'_1)^2+(u'_2)^2+(u'_3)^2+(u'_4)^2+(u'_5)^2 -
2D^2(E+ \frac{i\mu}{D^2}) + \frac{2D^4}{u^2_5}(E-\frac{i\mu}{D^2}) = 0,
$$
%=================================================================
\be
u''_j + 2(E+\frac{i\mu}{D^2}) u_j=0 ,
\quad j=1,2,3,4,
\label{3sphere}
\ee
%=================================================================
$$
u''_5+2(E+\frac{i\mu}{D^2})u_5 - \frac{2D^4}{u^3_5}
(E-\frac{i\mu}{D^2}) = 0,
$$
%=================================================================
subject to the constraints
%=================================================================
$$
\sum ^5_{k=1}u^2_k=D^2, \quad
\sum ^5_{k=1}u_ku'_k=0,
$$
%=================================================================
$$
\sum^5_{k=1}(u_ku''_k+(u'_k)^2) = 0,
\quad
u_4u'_3 - u_3u'_4 + u_2u'_1 - u_1u'_2 = 0.
$$
%=================================================================
Note that equations  (\ref{3sphere}) are compatible with these
constraints. Here, the Kepler problem on the sphere in three
dimensions is equivalent to choosing a Hamiltonian
%=================================================================
\bea
H&=&{1\over 2}(p^2_{u_1}+p^2_{u_2}+p^2_{u_3}+p^2_{u_4}+p^2_{u_5})
\\
&&
- (E+\frac{i\mu}{D^2})(u^2_1+u^2_2+u^2_3+u^2_4+u^2_5)
+\frac{D^4}{u^2_5}(E-\frac{i\mu}{D^2}),
\nonumber
\eea
%=================================================================
regarding the variables $u_j$ as independent and $\tau $ as time.
The only difference  is that there is now the constraint
%=================================================================
$$
u_4p_{u_3}-u_3p_{u_4}+u_2p_{u_1}-u_1p_{u_2}=0.
$$
%=================================================================

In terms of the Hamilton-Jacobi formulation we have the relation
%=================================================================
$$
\frac12(p^2_{s_1}+p^2_{s_2}+p^2_{s_3}+p^2_{s_4})
- \frac{\mu}{R}\frac{s_4}{\sqrt{s^2_1+s^2_2+s^2_3}} - E =
$$
%=================================================================
$$
- {u^2_5\over u^2_1+u^2_2+u^2_3+u^2_4}
\biggl[\frac{1}{2D^2}
(p^2_{u_1}+p^2_{u_2}+p^2_{u_3}+p^2_{u_4}+p^2_{u_5})
$$
%=================================================================
$$
- (E+ \frac{i\mu}{D^2})(u^2_1+u^2_2+u^2_3+u^2_4+u^2_5)
+ \frac{D^2}{u^2_5}(E-\frac{i\mu}{D^2})\biggr]=0.
$$
%=================================================================
With the usual substitutions, the corresponding Hamilton-Jacobi
equations are
%=================================================================
\be
\label{HJ1}
\sum_{k=1}^4
\left({\partial S\over \partial s_k}\right)^2
- \left(2E + \frac{2\mu}{R}\frac{s_4}{\sqrt{s_1^2 +s^2_2+s^2_3}}
\right)
=0,
\ee
%=================================================================
or
%=================================================================
\be
\label{HJ11}
\frac{1}{2D^2}
\sum_{k=1}^5
\left({\partial S\over \partial u_k}\right)^2
+ \left[\frac{D^2}{u^2_5}(E-\frac{i\mu}{D^2})
- (E+\frac{i\mu}{D^2})\right]
= 0,
\ee
%=================================================================
and the constraint has become
%=================================================================
\be
{\cal L} \cdot S = 0
\ee
%=================================================================
where operator ${\cal L}$ is
%=================================================================
\be
\label{CONSTR1}
{\cal L} =
u_2 {\partial \over \partial u_1}
- u_1 {\partial \over \partial u_2}
+ u_4 {\partial \over \partial u_3}
- u_3 {\partial \over \partial u_4}.
\ee
%=================================================================
Equation (\ref{HJ11}) can be solved by  separation of variables
in the spherical coordinates (\ref{COOR34}) on the complex sphere
$S_{4C}$.

\subsubsection{Quantum motion}

The associated quantum Kepler-Coulomb problem on the sphere
corresponding to the potential (\ref{POT2})
%=================================================================
\bea
\label{COUL34}
\frac{1}{2}\Delta_s^{(3)}\Psi + \left(E + \frac{\mu}{R}
\frac{s_4}{\sqrt{s^2_1+s^2_2+s^2_3}}\right)\Psi = 0
\eea
%=================================================================
translates directly to [see formula (\ref{LAP3})]
%=================================================================
\bea
\label{OS2}
\frac12\Delta_u^{(4)} \Phi
+\left({\cal E} - \frac{\omega^2 D^2}{2}
\frac{u_1^2+u_2^2+u_3^2+u_4^2}{u^2_5}\right)
\Phi = 0
\eea
%=================================================================
with the constraint
%=================================================================
\be
\label{CONS1}
{\cal L} \cdot \Phi = 0,
\ee
%=================================================================
where ${\cal L}$ is given by (\ref{CONSTR1}),
%=================================================================
\bea
\label{EX11}
\Psi = {u_5}^{\frac12} \Phi,
\eea
%=================================================================
and
%=================================================================
\bea
\label{EX1}
{\cal E} = 2i\mu - \frac{1}{D^2},
\qquad
\omega^2 D^2 = 2ED^2 - 2i\mu + \frac{3}{4D^2}.
\eea
%=================================================================
Here $\Delta_s^{(3)}$ and $\Delta_u^{(4)}$ are  Laplace-Beltrami
operators on the spheres S$_3$ and S$_{4C}$, respectively.

Consider the Schr\"odinger equation (\ref{OS2}) in complex
spherical coordinates (\ref{COOR34}). We have
%=================================================================
\bea
\label{RAD11}
\frac{{\mbox e}^{-i\chi}}{\sin^2\chi}\, \frac{\partial}{\partial\chi}
\,{\mbox e}^{i\chi}\sin^2\chi\, \frac{\partial \Phi}{\partial\chi}
+ \left[\omega^2 D^4 - i{\cal E}D^2 \frac{{\mbox e}^{i\chi}}{\sin\chi}
+\frac{{\vec L}^2}{\sin^2\chi}\right] \Phi = 0
\eea
%=================================================================
where the operator ${\vec L}^2$ are defined in (\ref{L1}).  We complexify the angle $\chi$ to the domain
{\bf G} (see Fig.1) by the transformation (\ref{COM}), such
that $\vartheta \in [0,\frac{\pi}{2}]$ and also complexify $\mu$
by setting $k=i\mu$ in expression for ${\cal E}$ and $\omega^2$.
Then equation (\ref{RAD11}) transforms to the Schr\"odinger equation
for the oscillator problem on real sphere S$_4$.

We make  the ansatz
%=================================================================
\bea
\label{WAVEF2}
\Phi(\vartheta,\alpha, \beta, \gamma) =
(\sin\vartheta)^{-\frac32}\,
Z(\vartheta) {\cal D}^{\ell}_{m_1, m_2}(\alpha, \beta, \gamma)
\eea
%=================================================================
where
%=================================================================
\bea
{\cal D}^{\ell}_{m_1, m_2}(\alpha, \beta, \gamma) =
{\mbox e}^{im_1\alpha}\,
d^{\ell}_{m_1, m_2}(\beta)\,
{\mbox e}^{im_2\gamma}
\eea
%=================================================================
is the Wigner function \cite{VAR}, satisfying the eigenvalue equation
%=================================================================
\bea
{\vec L}^2
{\cal D}^{\ell}_{m_1, m_2}(\alpha, \beta, \gamma) =
\ell(\ell+1) {\cal D}^{\ell}_{m_1, m_2}(\alpha, \beta, \gamma),
\eea
%=================================================================
and normalization condition
%=================================================================
\bea
\int {\cal D}^{\ell^{'*}}_{m_1', m_2'}(\alpha, \beta, \gamma)
{\cal D}^{\ell}_{m_1, m_2}(\alpha, \beta, \gamma)
\,
\frac18 \sin\beta d\beta d\alpha d\gamma
= \frac{2\pi^2}{2\ell+1}\,
\delta_{\ell \ell'} \delta_{m_1 m_1'} \delta_{m_2 m_2'}.
\eea
%=================================================================
Then the function $Z(\vartheta)$ satisfies
%=================================================================
\bea
\frac{d^2\, Z}{d \vartheta^2}
+ \left[\left(2{\cal E}D^2 + \omega^2 D^4 + \frac94 \right)
- \frac{\omega^2D^4}{\cos^2\vartheta} -
\frac{(2\ell+1)^2-\frac14}{\sin^2\vartheta}\right]
Z = 0.
\eea
%=================================================================
The corresponding solution regular at $\vartheta = 0, \pi/2$
and energy spectrum are given by
%================================================================
\bea
Z_{n_r \ell}(\vartheta)
&=& {\mbox{const}} \,
(\sin\vartheta)^{2\ell} (\cos\vartheta)^{\nu+\frac12}
\,
_2F_1(-n_r, \, n_r+2\ell+\nu+2; \, 2\ell+2; \, \sin^2\vartheta),
\\[2mm]
{\cal E}
&=& \frac{1}{2D^2}[(n+1)(n+4) + (2\nu-1)(n+2)],
\eea
%=================================================================
where $\nu = \left(\omega^2 D^4 + \frac14\right)^{\frac12}$,
$n=2n_r+2\ell=0,1,2..$ is the principal quantum number.  The
other quantum numbers are
%================================================================
$$
n_r=0,1,.. n, \qquad 2\ell= 0,1,... n, \qquad
m_1, m_2 = - \ell, -\ell+1, ...., \ell-1, \ell.
$$
%================================================================
Thus the wave function $\Phi(\vartheta,\alpha, \beta, \gamma)$
normalized under the condition (see Appendix)
%=================================================================
\bea
-\frac{iD^2}{2\pi}
\int_{S_{4C}} \Phi_{n_r\ell m_1 m_2} \,
\Phi^{\diamond}_{n_r\ell m_1 m_2}\,
{(u_1^2+u_2^2+u_3^2+u_4^2)}\, \frac{dv(u)}{u_5^2}
= 1
\eea
%=================================================================
has the form
%=================================================================
\bea
\label{WAVEF3}
\Phi_{n_r\ell m_1 m_2}(\vartheta,\alpha, \beta, \gamma)
=
C_{n_r \ell}(\nu)
\,
\sqrt{\frac{2\ell+1}{2\pi^2}}
\,
R_{n_r \ell}(\vartheta)
\,
{\cal D}^{\ell}_{m_1, m_2}(\alpha, \beta, \gamma)
\eea
%=================================================================
with
%================================================================
\bea
R_{n_r \ell}(\vartheta)
&=&
(\sin\vartheta)^{2\ell} (\cos\vartheta)^{\nu+\frac12}
\,
_2F_1(-n_r, \, n_r+2\ell+\nu+2; \, 2\ell+2; \, \sin^2\vartheta),
\\[2mm]
C_{n_r \ell}(\nu)
&=&
\frac{\sqrt{\pi}}{D^{\frac72}}\,
\sqrt{\frac{[(-i\nu)(\nu+2\ell+2n_r+2)]
(2\ell+n_r+1)!\Gamma(2\ell+\nu+n_r+2)}
{(1-e^{2i\pi\nu})
(\ell+n_r+1)[(2\ell+1)!]^2 (n_r)!\Gamma(\nu+n_r+1)}}.
\eea
%=================================================================

We now construct the wave function and energy spectrum for
the Schr\"odinger equation (\ref{COUL34}). The corresponding wave
function $\Psi(s)$ connecting with $\Phi(u)$ by  formula (\ref{EX11}) is independent of the variable $\gamma$ and $2\pi$ periodic in
$\alpha$ (the transformation $\alpha\rightarrow\alpha+2\pi$ is equivalent
the inversion $u_i\rightarrow -u_i$, $i=1,2,3,4$).
The constraint
(\ref{CONS1}) in the spherical coordinate (\ref{COOR34}) is
equivalent to
%=================================================================
$$
\frac{\partial}{\partial\gamma} \cdot
\Phi_{n_r \ell m_1 m_2}(\vartheta, \alpha, \beta, \gamma)
= m_2 \Phi_{n_r \ell m_1 m_2} (\vartheta, \alpha, \beta, \gamma)
= 0
$$
%=================================================================
and we have $m_2=0$. From $2\pi$ periodicity we get that $\ell$ and
$m_1$ are integers. Then, upon introducing the principal quantum number
$N = (n_r+\ell)+1 = \frac{n}{2}+1$ and using the expression
(\ref{EX1}), we  obtain the energy spectrum of
the reduced system
%=================================================================
\be
E =  \frac{N^2-1}{2R^2} - \frac{\mu^2}{2N^2},
\qquad N=1,2....
\ee
%=================================================================
where  $k=i\mu$. This  spectrum  coinsides
with that obtained from other methods \cite{SCHR,BIJ1,BOK}.

Returning from $\vartheta$ to the variable $\chi$, observing
that
%=================================================================
$$
\nu = i\sigma - N, \qquad \sigma = \frac{\mu R}{N}
$$
%=================================================================
and using the relations ($m_1\equiv m$)
%=================================================================
\be
\label{W1}
{\cal D}^{\ell}_{m, 0}(\alpha, \beta, \gamma)
=
(-1)^{m} \sqrt{\frac{4\pi}{2\ell+1}}
Y_{\ell m}(\beta, \alpha)
\ee
%=================================================================
we obtain the wave functions (with  correct normalization)
for the reduced system in the form
%=================================================================
$$
\Psi_{N \ell m}(\chi, \beta, \alpha)
=
\sqrt{D}\, e^{i\chi/2}\
\Phi_{n_r \ell m 0}(\chi, \alpha, \beta, \gamma)
=
\frac{(-1)^m}{\sqrt{R^3}} \,
{\mbox{\rm C}_{n_r \ell}(\sigma)}\,
\,
(\sin\alpha)^{\ell}
\,\,
{\mbox {\rm e}}^{-i\alpha(N-\ell-i\sigma)}
$$
\bea
\label{W2}
\times \,
_2F_1(-N+\ell+1, \, 1+\ell+i\sigma; \, 2\ell+2; \,
1-{\mbox {\rm e}}^{-2i\chi})
\,\,
Y_{lm}(\beta, \alpha),
\eea
%=================================================================
where
%=================================================================
$$
{\mbox{\rm C}_{n_r \ell}(\sigma)}\, =
\frac{2^{\ell+1}{\mbox {\rm e}}^{\frac{\pi \sigma}{2}}}{(2\ell+1)!}
\sqrt{\frac{(N^2+\sigma^2)(N+\ell)!}{2\pi N (N-\ell-1)!}}
|\Gamma(1+\ell+i\sigma)|.
$$
%=================================================================
This solution is identical to that given for the Coulomb eigenfunction
on S$_3$  in papers \cite{BOK,MPSV}.
Note that in  \cite{MPSV} it already has been shown that the function
(\ref{W2}) contracts as $R\rightarrow\infty$ into the flat space Coulomb wave
function for discrete and continuous energy spectrum.

\subsection{Generalized Hurwitz transformation}

The analogous problem in five dimensions can be realized via the variables
%=================================================================
\bea
\label{TRANS3}
s_1&=&\left(\sum_{k=1}^9u^2_k\right)^{\frac12}\,
\frac{i}{u_9} \, (u_1u_5+u_2u_6-u_3u_7-u_4u_8),\nonumber\\
s_2&=&\left(\sum_{k=1}^9u^2_k\right)^{\frac12}\,
\frac{i}{u_9}\, (u_1u_6-u_2u_5+u_3u_8-u_4u_7)\nonumber\\
s_3&=&\left(\sum_{k=1}^9u^2_k\right)^{\frac12}\,
\frac{i}{u_9}\, (u_1u_7+u_2u_8+u_3u_5+u_4u_6),\quad\\
s_4&=&\left(\sum_{k=1}^9u^2_k\right)^{\frac12}\,
\frac{i}{u_9}\, (u_1u_8-u_2u_7-u_3u_6+u_4u_5),\nonumber\\
s_5&=&\left(\sum_{k=1}^9u^2_k\right)^{\frac12}\,
\frac{i}{2u_9}\, (u^2_1+u^2_2+u^2_3+u^2_4-u^2_5-u^2_6-u^2_7-u^2_8),
\nonumber\\
s_6&=&\left(\sum_{k=1}^9u^2_k\right)^{\frac12}\,
\left(u_9+\frac{1}{2u_9}\sum_{k=1}^8u^2_k\right),
\nonumber
\eea
%=================================================================
which satisfy
%=================================================================
\be
\sum_{j=1}^6s^2_j = \left(\sum_{\ell=1}^9u^2_\ell\right)^2.
\ee
%=================================================================
The relation between the infinitesimal distances on the five dimensional
sphere S$_{5}$: \, $\sum_{i=1}^{6} s^2_i = R^2$
and eight dimensional complex sphere S$_{8C}$:
\, $\sum_{i=1}^{9} u^2_i = D^2$, ($R=D^2$) is
%=================================================================
\be
\frac{1}{R} \sum^6_{j=1}ds^2_j={-1\over u^2_9}
\left[(\sum^8_{k=1}u^2_k)\sum^9_{\ell =1}du
^2_\ell +\omega ^2_1+\omega ^2_2+\omega ^2_3\right],
\ee
%=================================================================
where
%=================================================================
\begin{eqnarray*}
\omega_1&=&u_4du_1+u_3du_2-u_2du_3-u_1du_4
-u_8du_5-u_7du_6+u_6du_7+u_5du_8,\\
\omega_2&=&u_3du_1-u_4du_2-u_1du_3
+u_2du_4-u_7du_5+u_8du_6+u_5du_7-u_6du_8,\\
\omega_3&=&u_2du_1-u_1du_2+u_4du_3
-u_3du_4+u_6du_5-u_5du_6+u_8du_7-u_7du_8,
\end{eqnarray*}
%=================================================================
and the constraint for mapping S$_{8C}\rightarrow$S$_5$ corresponds to
%=================================================================
\begin{eqnarray*}
\omega_i = 0, \qquad i=1,2,3.
\end{eqnarray*}
%=================================================================
Following \cite{HVT} (see also \cite{MST1}) we can supplement
the transformation (\ref{TRANS3}) with the angles
%=================================================================
\begin{eqnarray}
\label{DIRECT}
\alpha_H
&=&
\frac{1}{2}\left[{\mbox{arctan}}\frac{2u_1 u_2}{u_1^2-u_2^2}
+ {\mbox{arctan}}\frac{2u_3 u_4}{u_3^2-u_4^2}\right]
\,\, \in \, [0, 2\pi)
\nonumber\\[2mm]
\beta_H
&=& 2{\mbox{arctan}}
\left(\frac{u_3^2+u_4^2}{u_1^2+u_2^2}\right)^{\frac12}\,
\in \, [0, \pi]
\\[2mm]
\gamma_H
&=&
\frac{1}{2}\left[{\mbox{arctan}}\frac{2u_1 u_2}{u_1^2-u_2^2}
- {\mbox{arctan}}\frac{2u_3 u_4}{u_3^2-u_4^2}\right]
\,\, \in \, [0, 4\pi).
\nonumber
\end{eqnarray}
%=================================================================
The transformations (\ref{TRANS3}) and (\ref{DIRECT}) correspond to
 S$_{8C}$$\rightarrow$S$_8$=S$_5$$\otimes$S$_3$. If we now
choose the spherical coordinates on S$_{5}$ as
%=================================================================
\bea
\begin{array}{ll}
s_1 + is_2 = R \sin\chi \sin\vartheta
\cos{\frac{\beta}{2}} e^{i\frac{\alpha+\gamma}{2}},
&
s_5= R \sin\chi \cos\vartheta,
\\[2mm]
s_3 + is_4 = R \sin\chi \sin\vartheta
\sin{\frac{\beta}{2}} e^{i\frac{\alpha-\gamma}{2}},
&
s_6 = R \cos\chi.
\nonumber
\end{array}
\eea
%=================================================================
then the corresponding (nonorthogonal) spherical coordinates on
the eight dimensional complex sphere take the form ($D^2=R$)
%=================================================================
\begin{eqnarray}
\label{SPHER8}
u_1 &=& D\sqrt{1-e^{2i\chi}} \cos{\frac{\vartheta}{2}}
\cos{\frac{\beta_H}{2}} \cos{\frac{\alpha_H+\gamma_H}{2}}
\nonumber\\
u_2 &=& D\sqrt{1-e^{2i\chi}} \cos{\frac{\vartheta}{2}}
\cos{\frac{\beta_H}{2}} \sin{\frac{\alpha_H+\gamma_H}{2}}
\nonumber\\
u_3 &=&D\sqrt{1-e^{2i\chi}} \cos{\frac{\vartheta}{2}}
\sin{\frac{\beta_H}{2}} \cos{\frac{\alpha_H-\gamma_H}{2}}
\nonumber\\
u_3 &=&D\sqrt{1-e^{2i\chi}} \cos{\frac{\vartheta}{2}}
\sin{\frac{\beta_H}{2}} \sin{\frac{\alpha_H-\gamma_H}{2}}
\nonumber\\
u_5&=&D\sqrt{1-e^{2i\chi}} \sin{\frac{\vartheta}{2}}
\biggl(
\cos{\frac{\beta}{2}} \cos{\frac{\beta_H}{2}}
\cos{\frac{\alpha+\gamma+\alpha_H+\gamma_H}{2}}
\nonumber\\
&&+
\sin{\frac{\beta}{2}} \sin{\frac{\beta_H}{2}}
\cos{\frac{\alpha-\gamma-\alpha_H+\gamma_H}{2}}
\biggr)
\\
u_6&=&D\sqrt{1-e^{2i\chi}} \sin{\frac{\vartheta}{2}}
\biggl(
\cos{\frac{\beta}{2}} \cos{\frac{\beta_H}{2}}
\sin{\frac{\alpha+\gamma+\alpha_H+\gamma_H}{2}}
\nonumber\\
&&-
\sin{\frac{\beta}{2}} \sin{\frac{\beta_H}{2}}
\sin{\frac{\alpha-\gamma-\alpha_H+\gamma_H}{2}}
\biggr)
\nonumber\\
u_7&=&D\sqrt{1-e^{2i\chi}} \sin{\frac{\vartheta}{2}}
\bigl(
\sin{\frac{\beta}{2}} \cos{\frac{\beta_H}{2}}
\cos{\frac{\alpha-\gamma+\alpha_H+\gamma_H}{2}}
\nonumber\\
&&-
\cos{\frac{\beta}{2}} \sin{\frac{\beta_H}{2}}
\cos{\frac{\alpha+\gamma-\alpha_H+\gamma_H}{2}}
\bigr)
\nonumber\\
u_8&=&D\sqrt{1-e^{2i\chi}} \sin{\frac{\vartheta}{2}}
\bigl(
\sin{\frac{\beta}{2}} \cos{\frac{\beta_H}{2}}
\sin{\frac{\alpha-\gamma+\alpha_H+\gamma_H}{2}}
\nonumber\\
&&+
\cos{\frac{\beta}{2}} \sin{\frac{\beta_H}{2}}
\sin{\frac{\alpha+\gamma-\alpha_H+\gamma_H}{2}}
\bigr)
\nonumber\\
u_9 &=& D e^{i\chi}
\nonumber
\end{eqnarray}
%=================================================================
where $\chi\in [0,\pi]$, $\vartheta\in [0,\pi]$,
$\alpha \in [0, 2\pi]$, $\beta \in[0,\pi]$ and
$\gamma \in[0,4\pi]$.

\subsubsection{Classical motion}

The Kepler-Coulomb potential on the five dimensional sphere
${\mbox S_{5}}$ has the form
%=================================================================
\be
V= - \frac{\mu}{R}\, {s_6\over \sqrt{s^2_1+s^2_2+s^2_3+s^2_4+s^2_5}}.
\ee
%=================================================================
As before we can define a new coordinate $\tau $ such that
%=================================================================
$$
{d\tau \over dt}= \frac{1}{D^2}\, {u^2_9\over \sum ^8_{k=1}u^2_k}.
$$
%=================================================================
The corresponding equations of motion are given by
%=================================================================
\be
\sum ^9_{\ell =1}(u_\ell')^2 - 2(E+\frac{i\mu}{D^2})
- \frac{2D^2}{u^2_9} (E-\frac{i\mu}{D^2}) = 0,
\ee
$$
u''_k+2(E+\frac{i\mu}{D^2}) = 0, \quad k=1,\dots,8
$$
$$
u''_9+2(E+\frac{i\mu}{D^2})u_9
- \frac{2D^2}{u^3_9} (E-\frac{i\mu}{D^2}) = 0,
$$
%=================================================================
subject to the constraints
%=================================================================
\begin{eqnarray*}
\sum ^9_{\ell=1}u^2_\ell=D^2,    \quad
\sum ^9_{\ell=1}u_\ell u'_\ell=0,    \quad
\sum ^9_{\ell=1} \left(u_\ell {u''_\ell}+(u'_\ell)^2\right)&=&0,\\
u_4u'_1+u_3u'_2-u_2u'_3-u_1u'_4-u_8u'_5-u_7u'_6+u_6u'_7+u_5u'_8&=&0,\\
u_3u'_1-u_4u'_2-u_1u'_3+u_2u'_4-u_7u'_5+u_8u'_6+u_5u'_7-u_6u'_8&=&0,\\
u_2u'_1-u_1u'_2+u_4u'_3-u_3u'_4+u_6u'_5-u_5u'_6+u_8u'_7-u_7u'_8&=&0.
\end{eqnarray*}
%=================================================================

These equations of motion are equivalent to what we would obtain by
choosing the Hamiltonian
%=================================================================
\be
H = {1\over 2}\sum^9_{\ell=1} p^2_{u_\ell}
- (E+\frac{i\mu}{D^2})\sum^9_{\ell=1} u^2_\ell +
\frac{D^2}{u^2_9} (E-\frac{i\mu}{D^2}),
\ee
%=================================================================
regarding the variables $u_i$ as independent and using $\tau $ as time.
The associated constraints are
%=================================================================
\bea
u_4p_1+u_3p_2-u_2p_3-u_1p_4-u_8p_5-u_7p_6+u_6p_7+u_5p_8&=&0,
\nonumber\\
u_3p_1-u_4p_2-u_1p_3+u_2p_4-u_7p_5+u_8p_6+u_5p_7-u_6p_8&=&0,
\\
u_2p_1-u_1p_2+u_4p_3-u_3p_4+u_6p_5-u_5p_6+u_8p_7-u_7p_8&=&0.
\nonumber
\eea
%=================================================================

If we wish to solve this problem from the point of view of the
Hamilton-Jacobi equation we use the relation
%=================================================================
$$
\frac12\sum^6_{j=1}p^2_{s_j}
- \frac{\mu}{R^2} \frac{s_6}{\sqrt{s^2_1+s^2_2+s^2_3+s^2_4+s^2_5}}
- E=
$$
$$
-{u^2_9\over \sum ^8_{k=1}u^2_k}
\left\{\frac{1}{2D^2}\sum^9_{\ell=1} p^2_{u_\ell}
-(\frac{i\mu}{D^2}+E) \sum ^9_{\ell=1} u^2_\ell
+ \frac{D^2}{u^2_9} (E-\frac{i\mu}{D^2})\right\}=0.
$$
%=================================================================
The corresponding Hamilton-Jacobi equations are
%=================================================================
\be
\frac12\sum^6_{j=1}({\partial S\over \partial s_j})^2
- \frac{\mu}{R^2} \frac{s_6}{\sqrt{s^2_1+s^2_2+s^2_3+s^2_4+s^2_5}}-E=0,
\ee
\be
\frac{1}{2D^2} \sum^9_{\ell=1}({\partial S\over \partial u_\ell})^2
-(\frac{i\mu}{D^2}+E)\sum^9_{\ell=1}u^2_\ell +
\frac{D^2}{u^2_9} (E-\frac{i\mu}{D^2}) = 0,
\ee
%=================================================================
subject to the constraints
%================================================================
\begin{eqnarray*}
u_4 {\partial S\over \partial u_1}+u_3 {\partial S\over \partial u_2}-u_2
{\partial S\over \partial u_3}-u_1 {\partial S\over \partial u_4}-u_8
{\partial S\over \partial u_5}-u_7 {\partial S\over \partial u_6}+u_6
{\partial S\over \partial u_7}+u_5 {\partial S\over \partial u_8}
&=& 0,
\\
u_3 {\partial S\over \partial u_1}-u_4 {\partial S\over \partial u_2}-u_1
{\partial S\over \partial u_3}+u_2 {\partial S\over \partial u_4}-u_7
{\partial S\over \partial u_5}+u_8 {\partial S\over \partial u_6}+u_5
{\partial S\over \partial u_7}-u_6 {\partial S\over \partial u_8}
&=& 0, \\
u_2 {\partial S\over \partial u_1}-u_1 {\partial S\over \partial u_2}+u_4
{\partial S\over \partial u_3}-u_3 {\partial S\over \partial u_4}+u_6
{\partial S\over \partial u_5}-u_5 {\partial S\over \partial u_6}+u_8
{\partial S\over \partial u_7}-u_7 {\partial S\over \partial u_8}
&=& 0.
\end{eqnarray*}
%=================================================================

\subsubsection{Quantum motion}

The Schr\"odinger equation for the 5-dimensional quantum
Coulomb problem
%=================================================================
\bea
\label{COULM5}
\frac12\Delta_s^{(5)}\Psi +
\left(E+  \frac{\mu}{R}
\frac{s_6}{\sqrt{s^2_1+s^2_2+s^2_3+s^2_4+s^2_5}}
\right)\Psi = 0
\eea
%==================================================================
transforms to the 8-dimensional oscillator equation (see
Appendix)
%=================================================================
\bea
\label{OSCIL8}
\frac12\Delta_u^{(8)} \Phi +
\left({\cal E}-
\frac{\omega^2D^2}{2}
\frac{1}{u_9^2}
\sum_{i=1}^{8}u_i^2 \, \right)\Phi = 0
\eea
%=================================================================
with constraints
%=================================================================
\bea
T_i \Phi = 0,
\eea
%================================================================
where operator ${\vec T}$ is given by formula (\ref{T}),
%=================================================================
\bea
{\cal E} = \left(2i\mu-\frac{6}{D^2}\right),
\qquad
\omega^2D^2 = 2\left(D^2 E - 2i\mu + \frac{15}{8D^2}\right),
\eea
%=================================================================
and
%=================================================================
\bea
\Psi = (u_9)^{\frac32} \, \Phi.
\eea
%=================================================================
Considering the oscillator equation (\ref{OSCIL8}) in  complex
spherical coordinates (\ref{SPHER8}) we get (see Appendix)
%=================================================================
\bea
\label{OSCIL81}
\frac{{\mbox e}^{-3i\chi}}{\sin^4\chi}\,
\frac{\partial}{\partial\chi}
\,
{\mbox e}^{3i\chi}\sin^4\chi\, \frac{\partial \Phi}{\partial\chi}
+ \left[\omega^2 D^4 - i{\cal E}D^2 \frac{{\mbox e}^{i\chi}}{\sin\chi}
+\frac{{\vec M}^2}{\sin^2\chi}\right] \Phi = 0,
\eea
%=================================================================
where the operator ${\vec M}^2$ has the form
%=================================================================
\bea
{\vec M}^2 =
\frac{1}{\sin^3\theta}\,
\frac{\partial}{\partial\theta}
\,
\sin^3\theta\,
\frac{\partial}{\partial\theta} -
\frac{{\vec L}^2}{\sin^2\frac{\theta}{2}}-
\frac{{\vec J}^2}{\cos^2\frac{\theta}{2}},
\eea
%=================================================================
and
%=================================================================
\bea
{\vec J} = {\vec L} + {\vec T},
\qquad
{\vec J}^2 = {\vec L}^2 + {\vec T}^2 + 2 {\vec L}\cdot {\vec T}.
\eea
%=================================================================
As before, we make the complex transformation (\ref{COM})
and also complexify parameter $\mu$ by putting $k=i\mu$. We make
the separation ansatz \cite{MST1}
%=================================================================
\bea
\label{OSCIL82}
\Phi = R (\vartheta) Z(\theta) G(\alpha, \beta, \gamma;
\alpha_H, \beta_H, \gamma_H)
\eea
%=================================================================
where  $G$ is an eigenfunction of operators
${\vec L}^2$, ${\vec T}^2$ and ${\vec J}^2$ with  eigenvalues
$L(L+1)$, $T(T+1)$,  $J(J+1)$, respectively. Correspondingly the
wave function $Z(\theta)$ is the eigenfunction of operator
${\vec M}^2$ with eigenvalue $\lambda(\lambda+3)$.
Because there is ${\vec L}\cdot{\vec T}$ interaction the eigenvalue
equation
%=================================================================
\bea
\label{OSCIL83}
{\vec J}^2 G(\alpha, \beta, \gamma; \alpha_H, \beta_H, \gamma_H)
=
J(J+1) G(\alpha, \beta, \gamma; \alpha_H, \beta_H, \gamma_H)
\eea
%=================================================================
can not be separated in variables $(\alpha, \beta, \gamma;
\alpha_H, \beta_H, \gamma_H)$ but we can apply the rules for the
addition of angular momenta ${\vec L}$ and ${\vec T}$ and, following
\cite{MST1} express $G$ as a
 Clebsch-Gordan expansion
%=================================================================
\bea
\label{OSCIL84}
G^{JM}_{Lm;Tt}
=
\sum_{M=m'+t'}\, (J,M|L,m'; \, T,t')
\,\,
{\cal D}^{L}_{m, m'}(\alpha, \beta, \gamma)
\,
{\cal D}^{T}_{t, t'}(\alpha_H, \beta_H, \gamma_H)
\eea
%=================================================================
where $(JM|Lm;Tt)$ are the Clebsch-Gordan coefficients. Note
that the functions $G^{JM}_{Lm;Tt}$ satisfy the normalization
condition
%================================================================
\bea
\int_{\Omega }d\Omega \int_{\Omega_H} d\Omega_H
G^{JM}_{Lm;Tt} G^{{J'M'}*}_{L'm';T't'} \,
= \left(\frac{2\pi^2}{2L+1}\right) \,
\left(\frac{2\pi^2}{2T+1}\right) \,
\delta_{JJ'}\delta_{LL'}\delta_{TT'}\delta_{MM"}
\delta_{mm'}\delta_{tt'}.
\eea
%=================================================================
If we substitute ansatz (\ref{OSCIL82}) into the Schr\"odinger
equation (\ref{OSCIL81}), then after separation of variables we
obtain the  differential equations
%=================================================================
\bea
\label{OSCIL85}
\frac{1}{\sin^3\theta}\,
\frac{d}{d\theta}
\,
\sin^3\theta\,
\frac{d Z}{d\theta} +
\left[\lambda(\lambda+3) -
\frac{2L(L+1)}{1-\cos\theta}-
\frac{2J(J+1)}{1+\cos\theta}
\right] Z = 0,
\eea
%=================================================================
%=================================================================
\bea
\label{OSCIL86}
\frac{1}{\sin^7\vartheta}\,
\frac{d}{d\vartheta}
\,
\sin^7\theta\,
\frac{d R}{d\vartheta} +
\left[(2D^2{\cal E}+\omega^2D^4) -
\frac{4\lambda(\lambda+3)}{\sin^2\vartheta}-
\frac{\omega^2 D^4}{\cos^2\vartheta}\right]=0,
\eea
%================================================================
with  real parameters
%================================================================
\bea
{\cal E} = \left(2k-\frac{6}{D^2}\right),
\qquad
\omega^2D^2 = 2\left(D^2 E - 2k + \frac{15}{8D^2}\right).
\eea
%=================================================================

Consider equation (\ref{OSCIL85}). Taking the
new function by $v(\theta) = (\sin\theta)^{\frac32} Z(\theta)$
we obtain the P\"oschl-Teller equation. Then the solution
$Z(\theta)\equiv Z^{JL}_{\lambda}(\theta)$ orthonormalized by the
condition
%================================================================
\bea
\int_{0}^{\pi} Z^{JL}_{\lambda}(\theta)
Z^{JL*}_{\lambda'}(\theta)\, \sin^3\theta \, d\theta
= \delta_{\lambda \lambda'}
\eea
%=================================================================
has the form
%================================================================
\bea
\label{OSCIL87}
Z^{JL}_{\lambda}(\theta) =
\sqrt{\frac{(2\lambda+3)(\lambda+J+L+2)!
(\lambda-L-J)!}
{2^{2J+2L+2}(\lambda-L+J+1)!(\lambda-J+L+1)!}}
\nonumber\\[2mm]
\times
(1-\cos\theta)^{J} \,
(1+\cos\theta)^{L} \,
P^{(2L+1, 2J+1)}_{n_\theta}(\cos\theta),
\qquad
n_\theta=0,1,2,...
\eea
%=================================================================
where $\lambda$ is quantized as $\lambda-L-J=n_\theta$.

Let us now turn to the quasiradial equation (\ref{OSCIL86}). Setting
 $w(\vartheta)= (\sin\vartheta)^{-\frac72}R(\vartheta)$,
we can rewrite this equation in the P\"oschl-Teller form
%=================================================================
\bea
\label{OSCIL88}
\frac{d^2 w}{d\vartheta^2} +
\left[(2D^2{\cal E}+\omega^2D^4+\frac{49}{4}) -
\frac{(2\lambda+3)^2-\frac14}{\sin^2\vartheta}-
\frac{\omega^2D^4}{\cos^2\vartheta}\right] w = 0.
\eea
%================================================================
Solving this equation we have following expression for
quasiradial functions $R(\vartheta) \equiv R_{n_r\lambda}(\theta)$:
%================================================================
\bea
\label{OSCIL89}
R_{n_r\lambda}(\theta)&=&(\sin\vartheta)^{2\lambda} \,
(\cos\vartheta)^{\nu+\frac12}, \,
\nonumber\\[2mm]
&\times&
{_2F_1}(-n_r, \, n_r+\nu+2\lambda+4; \, 2\lambda+4; \,
\sin^2\vartheta),
\,\,\,\,
n_r =0,1,2,...
\eea
%=================================================================
with energy levels given by
%================================================================
\bea
\label{OSCIL90}
{\cal E} = \frac{1}{2D^2}[(n+1)(n+8) + (2\nu-1)(n+4)],
\qquad n=0,1,2...
\eea
%=================================================================
where $\nu = \left(\omega^2 D^4 + \frac14\right)^{\frac12}$,
and  principal quantum number
%================================================================
$$
n=2(n_r+\lambda) = 2(n_r+n_\theta+L+J).
$$
%================================================================

Thus, the full wave function $\Phi$ is the simultaneous eigenfunction
of the Hamiltonian and commuting operators $M^2, {\vec J}^2, {\vec L}^2,
{\vec T}^2, {J_3}, {L_3}$ and ${T_3}$. The explicit form of this
function satisfying  the normalization condition (see Appendix)
%=================================================================
$$
- \frac{iD^5}{32\pi^2}
\int_{S_{8C}} \,
\Phi^{JLT}_{n_r \lambda M m t}\,
\Phi^{JLT\, \diamond}_{n_r \lambda M m t}\,
\sum_{i=1}^{8}{u_i^2}\,
\frac{dv(u)}{u_9^2} = 1
$$
%=================================================================
is
%================================================================
\bea
\label{OSCIL91}
\Phi^{JLT}_{n_r \lambda M m t}
=
C_{n_r \lambda}(\nu)\,
\frac{\sqrt{(2L+1)(2T+1)}}{2\pi^2}\,
R_{n_r\lambda}(\vartheta)\,
Z^{JL}_{\lambda}(\theta)\,
G^{JM}_{Lm;Tt}(\alpha, \beta, \gamma; \alpha_H, \beta_H, \gamma_H)
\eea
%=================================================================
where $R_{n_r\lambda}(\vartheta)$ is given by formula (\ref{OSCIL89})
and
%================================================================
\bea
\label{OSCIL92}
C_{n_r \lambda}(\nu) =
\frac{4}{(2\lambda+3)!}\,
\sqrt{\frac{i\nu(\nu+2\lambda+2n_r+4)\Gamma(2\lambda+\nu+n_r+4)
(n_r+2\lambda+3)!}
{D^{13} \pi^2 (1-e^{2i\pi\nu})(\lambda+n_r+2)(n_r)!
\Gamma(\nu+n_r+1)}}.
\eea
%=================================================================

Let us construct now the five-dimensional Coulomb system.
The  constraints tell us
%================================================================
\bea
{\vec T}^2 \Phi (u) = T(T+1)\Phi(u) = 0.
\eea
%================================================================
and therefore the oscillator eigenstates span  the states with
$T=0$ and $L=J$. For $L=J$ the Jacobi polynomial in (\ref{OSCIL87})
is proportional to the Gegenbauer polynomial \cite{BE}
%================================================================
\bea
P^{(2L+1, 2L+1)}_{\lambda-2L}(\cos\theta)
=
\frac{(4L+2)!(\lambda+1)!}{(2l+1)!(2L+\lambda+2)!}\,\,
C^{2L+\frac32}_{\lambda-2L}(\cos\theta),
\eea
%=================================================================
and we obtain
%================================================================
\bea
\label{OSCIL871}
Z^{JL}_{\lambda}(\theta)
&\equiv&
Z_{l\lambda}(\theta)
\nonumber\\[2mm]
&=&
2^{2L+1} \,
\Gamma(2L+\frac32)\,
\sqrt{\frac{(2\lambda+3)(\lambda-2L)!}
{\pi (\lambda+2L+2)!}}
\,
(\sin\theta)^{2L} \,\,
C^{2L+\frac32}_{\lambda-2L}(\cos\theta).
\eea
%=================================================================
Then from properties of Clebsch-Gordan coefficients $(JM|Lm;00) =
\delta_{JL} \delta_{Mm'}$ and using
${\cal D}^{0}_{0, 0}(\alpha_H, \beta_H, \gamma_H)= 1$ we see that
the expansion (\ref{OSCIL84}) yields
%=================================================================
\bea
\label{OSCIL93}
G^{JM}_{Lm;00}(\alpha, \beta, \gamma; \alpha_H, \beta_H, \gamma_H)
=
{\cal D}^{L}_{m, m'}(\alpha, \beta, \gamma)
\,
\delta_{JL} \, \delta_{Mm'}.
\eea
%=================================================================
Thus, the function $\Phi$ now depends only on variables
($\vartheta,\theta,\alpha,\beta,\gamma$). Observing that
$\lambda=n_\theta+2L = 0,1,2,...n$,  introducing the new principal
quantum number $N=(n_r+\lambda)= \frac{n}{2} = 0,1,2..$ and setting  $k=i\mu$, we easily get
>from the oscillator energy spectrum (\ref{OSCIL90}) the reduced system energy levels
%================================================================
\bea
\label{OSCIL94}
E_{N} = \frac{N(N+4)}{2R^2}-\frac{\mu^2}{2(N+2)^2}.
\eea
%=================================================================
Noting that $\nu=i\sigma-(N+2)$ and taking into account the formulas
(\ref{OSCIL89}) and (\ref{OSCIL91})-(\ref{OSCIL93}), we finally have
the solution of the Schr\"odinger equation (\ref{COULM5}) as
%=================================================================
\bea
\label{OSCIL95}
\Psi^{L\lambda}_{n_r m m'}(\chi,\theta; \alpha, \beta, \gamma)
&=&
D^{\frac32}\, e^{\frac32 i\chi}
\,
\Phi^{L\lambda}_{n_r mm'}(\chi,\theta; \alpha, \beta, \gamma)
\nonumber\\[2mm]
&=&
N^{L\lambda}_{n_r}(\sigma)\,
R_{n_r\lambda}(\vartheta)\,
Z_{L\lambda}(\theta)\,
\,
\sqrt{\frac{2L+1}{2\pi^2}}\,
{\cal D}^{L}_{m, m'}(\alpha, \beta, \gamma)
\eea
%=================================================================
where  $Z_{L\lambda}(\theta)$ is given by (\ref{OSCIL871})
and
%=================================================================
\bea
R_{n_r\lambda}(\chi)
&=&
(\sin\chi)^{\lambda} \,
e^{-i\chi(N-\lambda-i\sigma)}
\,
{_2F_1}(-N+\lambda, \, \lambda+2+i\sigma; \, 2\lambda+4; \,
1-e^{2i\chi}),
\\[2mm]
N^{L}_{n_r \lambda}(\sigma)
&=&
\label{COUL8}
\frac{2^{\lambda+2} e^{\frac{\pi\sigma}{2}}}{(2\lambda+3)!}
\,
\sqrt{\frac{[(N+2)^2+\sigma^2](N+\lambda+3)!}
{2 R^5 \pi(N+2)(N-\lambda)!}}
\,
|\Gamma(\lambda+2+i\sigma)|.
\eea
%=================================================================
Thus, we have constructed the wave function and energy spectrum for the
five-dimensional Coulomb problem.
In the contraction limit $R\rightarrow\infty$ for finite $N$
we get the formula for the discrete energy spectrum of the
five-dimensional Coulomb problem \cite{KMT}
%=================================================================
$$
\lim_{R\rightarrow\infty} E_N(R) = - \frac{\mu^2}{(N+2)^2},
\qquad N=0,1,...
$$
%=================================================================
Taking the limit $R\rightarrow\infty$ and using  asymptotic
formulas as in (\ref{CONTR1}) we get from (\ref{OSCIL95})-(\ref{COUL8})
%=================================================================
\bea
\lim_{R\rightarrow\infty}\,
\Psi^{L\lambda}_{n_r m m'}(\chi,\theta; \alpha, \beta, \gamma)
&=&
R_{N\lambda}(r)\, Z_{L\lambda}(\theta)\,
\,
\sqrt{\frac{2L+1}{2\pi^2}}\,
{\cal D}^{L}_{m, m'}(\alpha, \beta, \gamma)
\eea
%=================================================================
with
%=================================================================
$$
R_{N\lambda}(r)\,
= \frac{4\mu^{5/2}}{(N+2)^3} \,
\sqrt{\frac{(N+\lambda+3)!}{(N-\lambda)!}}
\left(\frac{2\mu r}{N+2}\right)^{\lambda}\,
\frac{e^{-\frac{\mu r}{N+2}}}
{(2\lambda+3)!}\,
{_1F_1}(-N+\lambda; \, 2\lambda+4; \, \frac{2\mu r}{N+2}),
$$
%=================================================================
which coincides with the five-dimensional Coulomb wave function
obtained in paper \cite{KMT}.

\section{Summary and Discussion}

In this paper we
have constructed a series of mappings S$_{2C}\rightarrow ${S}$_2$,
S$_{4C}\rightarrow ${S}$_3$ and S$_{8C}\rightarrow ${S}$_5$, that
are generalize those well known from the  Euclidean space
Levi-Civita, Kustaanheimo-Steifel and Hurwitz transformations.
We have shown, that as in case of flat space, these
transformations permit one to establish the {\it correspondence}
between the
Kepler-Coulomb and oscillator problems in classical and quantum
mechanics for the respective dimensions. We have seen that using these
generalized transformations (\ref{TR1}), (\ref{KS}) and (\ref{TRANS3})
we can completely solve the quantum Coulomb system on the two-, three- and
five-dimensional sphere, including eigenfunctions with correct
normalization constant and energy spectrum.

For the solution of the  quantum Coulomb problem,  first we transformed
the Schr\"odinger equation to the equation with oscillator potential
on the complex sphere. Then, via  complexification of the
Coulomb coupling constant $\mu$ ($\mu=Ze^2$) and the quasiradial
variable $\chi$ this problem was translated to the oscillator system
on the real sphere and  solved.

It is interesting to note that the  complexification of
constant $Ze^2/R$ and the quasiradial variable were first used
by Barut, Inomata and Junker \cite{BIJ1} in the path integral approach
to the Coulomb system on the three-dimensional sphere and hyperboloid, and
further were applied to  two- and three- dimensional superintegrable
systems on spaces with constant curvature \cite{GRO1,GRO2}. The
substitution used in \cite{BIJ1}
%================================================================
\bea
\label{BAR}
{\mbox e}^{i\chi} = - \coth \beta,
\qquad
\beta\in (-\infty, \infty)
\eea
%=================================================================
is correct as an analytic continuation to the region $0\leq {\mbox Re}
\chi \leq \pi$ and $-\infty < {\mbox Im} \leq 0$ and translates the
Coulomb quasiradial equation with variable $\chi$ to the modified
P\"oschl-Teller equation with variable $\beta$. It is possible to show
that there exists a connection between (\ref{BAR}) and generalized
Levi-Civita transformations on  constant curvature spaces.
Indeed, for instance, along with the mapping S$_{2C}$$\rightarrow$S$_2$
we can determine a mapping H$_{2C}$$\rightarrow$S$_2$, i.e. from the
two-dimensional complex hyperboloid to the real sphere:
%=================================================================
$$
s_1^2+s_2^2+s_3^2 = (u^2_3 - u^2_1 - u^2_2)^2.
$$
%=================================================================
This transformation has the form
%=================================================================
\bea
\label{TRH2}
s_1 &=& i \sqrt{u^2_3 - u^2_1 - u^2_2}
\cdot\frac{u^2_1-u^2_2}{2u_3},
\nonumber\\
s_2 &=& i \sqrt{u^2_3 - u^2_1 - u^2_2}
\cdot\frac{u_1u_2}{u_3},
\\
s_3 &=&  \sqrt{u^2_3 - u^2_1 - u^2_2}
\cdot\left(u_3 - \frac{u^2_1+u^2_2}{2u_3}\right),
\nonumber
\eea
%=================================================================
and translates the Schr\"odinger equation for the Coulomb problem on the
sphere to the oscillator problem on the complex hyperboloid.
Then the substitution (\ref{BAR}) transforms the oscillator problem
>from the complex to the real hyperbolid, a solution well known from
papers \cite{GRO2,KMJP2}.

The method described in this paper can be applied not just to (\ref{coulombpot2}) but to many
Coulomb-like potentials. In particular the generalized two-dimensional
Kepler-Coulomb problem may be transformed to the Rosokhatius system on
the
two-dimensional sphere \cite{KMJP}.

As we have seen, in spite of the similarity of transformations
(\ref{MAT1}) and (\ref{TR1}) on the sphere and Euclidean space
there exist  essential differences. Equations (\ref{TR1}), (\ref{KS})
and (\ref{TRANS3}) determine the transformations between complex and
real spheres or in ambient spaces a mapping $C_{2p+1}\rightarrow
R_{p+2}$ for $p=1,2,4$. Evidently these facts are closely connected
to  Hurwitz theorem \cite{HUR}, according to which
the nonbijective bilinear transformations satisfy the identity
%=================================================================
\bea
\label{IDEN}
s_1^2+s_2^2+ .... + s_f^2 = (u^2_1+u^2_2+ ... + u^2_n)^2
\eea
%=================================================================
 only for four pair of dimensions: $(f,n) = (1,1), (2,2), (3,4)$
and $(5,8)$, which corresponds to a mapping
$R_{2p}\rightarrow R_{p+1}$ for $p=1,2,4$ respectively.

For transformations between real spaces of constant curvature the
situation is more complicated, and more interesting. For example,
the two-dimensional transformation on the hyperboloid is
%=================================================================
\bea
\label{TRH11}
s_1 &=&  \sqrt{u^2_3 \pm u^2_1 \pm u^2_2}
\cdot\frac{u^2_1-u^2_2}{2u_3},
\nonumber\\
s_2 &=&  \sqrt{u^2_3 \pm u^2_1 \pm u^2_2}
\cdot\frac{u_1u_2}{u_3},
\\
s_3 &=&  \sqrt{u^2_3 \pm u^2_1 \pm u^2_2}
\cdot\left(u_3 \pm \frac{u^2_1+u^2_2}{2u_3}\right),
\nonumber
\eea
%=================================================================
and
%=================================================================
\bea
s_3^2-s_1^2-s_2^2 = (u^2_3 \pm u^2_1 \pm u^2_2)^2.
\eea
%=================================================================
Thus, the upper and lower hemispheres of the real sphere or
the upper and lower sheets of the two-sheet hyperboloid in
$u$-space map to the upper and lower sheets, respectively,
of the two-sheet hyperboloid in $s$-space.

The next example is the transformation
%=================================================================
\bea
\label{TRH22}
s_1 &=&  \sqrt{u^2_1 + u^2_2 - u^2_3}
\cdot\frac{u^2_1-u^2_2}{2u_3},
\nonumber\\
s_2 &=&  \sqrt{u^2_1 + u^2_2 - u^2_3}
\cdot\frac{u_1u_2}{u_3},
\\
s_3 &=&  \sqrt{u^2_1 + u^2_2 - u^2_3}
\cdot\left(u_3 - \frac{u^2_1+u^2_2}{2u_3}\right),
\nonumber
\eea
%=================================================================
and
%=================================================================
\bea
s_1^2+s_2^2-s_3^2 = (u^2_1 + u^2_2 - u^2_3)^2.
\eea
%=================================================================
Here the one-sheet hyperboloid in $u$-space maps to the one-sheet
hyperboloid in $s$-space. From transformations (\ref{TRH11}) and
(\ref{TRH22}) (using methods as in \S2) it is
easy to show that in the contraction limit $D\rightarrow\infty$
this transformation  goes to the real Levi-Civita transformation
(up to the translation ${\bar u}_i \rightarrow \sqrt{2}{\bar u}_i$)
(\ref{MAT1}).This shows that the method of this article can be adapted
to treat a
Kepler-Coulomb system on the two- and one sheet hyperboloids.

Finally, note that in this article we do not discuss two important
questions. First  is the correspondence between integrals of motion
for Kepler-Coulomb and oscillator systems. Second is the connection
between separable systems of coordinates (not only spherical) under
mappings (\ref{TR1}), (\ref{KS}) and (\ref{TRANS3}).
This investigation
will be carried out elsewhere.

\section{Appendix}

We present some differential aspects of
the generalized Levi-Civita, KS and Hurwitz transformations.
These calculations we are related to those in
\cite{KIB3,MST1} for flat space.

\vspace{0.6cm}
\noindent
5.1 {\bf Transformation} S$_{2C}\rightarrow ${S}$_2$

The Laplace-Beltrami operator on the $u$ - sphere in
complex spherical coordinates (\ref{CU}) is
%=================================================================
\bea
\Delta_u^{(2)} &=& \frac{1}{D^2}
\left[(u_1\partial_{u_2} - u_2\partial_{u_1})^2 +
(u_3\partial_{u_2} - u_2\partial_{u_3})^2 +
(u_3\partial_{u_2} - u_2\partial _{u_3})^2\right]
\nonumber\\
&=& \frac{2i}{D^2} \sin\chi {\mbox e^{- i\chi}}
\left\{\frac{1}{\sin\chi}\frac{\partial}{\partial\chi}
\sin\chi
\frac{\partial}{\partial\chi} + \frac{1}{\sin^2\chi}
\frac{\partial^2}{\partial\varphi^2}\right\}
\eea
%=================================================================
while the usual Laplace-Beltrami operator on the $s$ - sphere in
spherical coordinates  ($\chi, \varphi$) has the form
%=================================================================
\bea
\Delta_s^{(2)}
&=& \frac{1}{R^2}\left[
(s_1\partial_{s_2} - s_2\partial_{s_1})^2 +
(s_3\partial_{s_2} - s_2\partial_{s_3})^2 +
(s_3\partial_{s_2} - s_2\partial_{s_3})^2\right]
\nonumber\\
&=&\frac{1}{R^2}
\left\{\frac{1}{\sin\chi}\frac{\partial}{\partial\chi}
\sin\chi
\frac{\partial}{\partial\chi} + \frac{1}{\sin^2\chi}
\frac{\partial^2}{\partial\varphi^2}\right\}
\eea
%=================================================================
The two Laplacians are connected through
%=================================================================
\bea
\label{LB22}
\Delta_s^{(2)} = - \frac{u^2_3}{u^2_1+u^2_2} \,
\frac{1}{D^2} \, \Delta_{u}^{(2)}.
\eea
%=================================================================
The volume elements in $u$ and $s$ - spaces are
%=================================================================
\bea
dv(u) = - \frac{iD^2}{2}{{\mbox e}^{i\chi}} d\chi d\varphi,
\qquad
dv(s) = R^2 \sin\chi d\chi d\varphi
\eea
%=================================================================
and
%=================================================================
\bea
\label{VOLUM1}
\frac{1}{R}\, d v(s) =
- \frac{u^2_1+u^2_2}{u^2_3} \,   dv(u).
\eea
%=================================================================
We have (the variable $\varphi$ runs the  from
$0$ to $4\pi$)
%=================================================================
\bea
\label{APINT2}
\int_{S_{2}}\, .... \, dv(s) =
- \frac{D^2}{2} \int_{S_{2C}}\, ....
\frac{u^2_1+u^2_2}{u^2_3} \, d v(u).
\eea
%=================================================================

\vspace{0.6cm}
\noindent
5.2 {\bf Transformation}
${\mbox S_{4{\bf c}}} \rightarrow {\mbox S_{3}}$

The Laplace-Beltrami operator on the $u$ - sphere in
($\chi, \alpha, \beta, \gamma$) coordinates is
%=================================================================
\bea
\Delta_u^{(4)}
= \frac{2i}{D^2} \sin\chi {\mbox e^{- i\chi}}
\left[\frac{{\mbox e^{- i\chi}}}
{\sin^2\chi}\frac{\partial}{\partial\chi}
{\mbox e^{i\chi}}\sin^2\chi\frac{\partial}{\partial\chi}
+ \frac{{\vec L}^2}{\sin^2\chi}\right]
\eea
%=================================================================
where
%=================================================================
\bea
\label{L1}
{L}_1 &=&
i\left(\cos\alpha\cot\beta\frac{\partial}{\partial\alpha}
+ \sin\alpha\frac{\partial}{\partial\beta} -
\frac{\cos\alpha}{\sin\beta}\frac{\partial}{\partial\gamma}\right),
\nonumber\\[2mm]
{L}_2 &=&
i\left(\sin\alpha\cot\beta\frac{\partial}{\partial\alpha}
- \cos\alpha\frac{\partial}{\partial\beta} -
\frac{\sin\alpha}{\sin\beta}\frac{\partial}{\partial\gamma}\right),
\\[2mm]
{L}_3 &=&
- i\frac{\partial}{\partial\alpha},
\nonumber
\eea
%=================================================================
and
%=================================================================
\bea
{\vec L}^2 = \biggl[\frac{\partial^2}{\partial\beta^2}
+ \cot\beta\frac{\partial}{\partial\beta}
+ \frac{1}{\sin^2\beta}
\biggl(\frac{\partial^2}{\partial\gamma^2}
- 2\cos\beta\frac{\partial}{\partial\gamma}
\frac{\partial}{\partial\alpha}
+ \frac{\partial^2}{\partial\alpha^2}\biggr)\biggr],
\eea
%=================================================================
while the usual Laplace-Beltrami operator on the $s$ - sphere in
($\chi, \beta, \alpha$) coordinates is
%=================================================================
$$
\Delta_s^{(3)}
= \frac{1}{R^2}
\left[\frac{1}{\sin^2\chi}\frac{\partial}{\partial\chi}
\sin^2\chi\frac{\partial}{\partial\chi}
+ \frac{1}{\sin^2\chi}
\left(\frac{\partial^2}{\partial\beta^2}
+ \cot\beta\frac{\partial}{\partial\beta}
+ \frac{1}{\sin^2\beta}
\frac{\partial^2}{\partial\alpha^2}\right)\right].
$$
%=================================================================
The two Laplace-Beltrami operators are connected by
%=================================================================
$$
\Delta_u^{(4)}
= \frac{2i}{D^2} \sin\chi
{\mbox e^{- \frac{3i}{2}\chi}}
\biggl[D^4 \Delta_s^{(3)}
+ \left(\frac14- i\cot\chi\right)
+ \frac{1}{\sin^2\chi}\frac{1}{\sin^2\beta}
$$
$$
\times \frac{\partial}{\partial\gamma}
\left(\frac{\partial}{\partial\gamma}
- 2\cos\beta\frac{\partial}{\partial\alpha}\right)\biggr]
\, {\mbox e^{\frac{i}{2}\beta}},
$$
%=================================================================
and  the operator acting on  functions of variables
($\chi, \beta, \alpha$) is
%=================================================================
\bea
\label{LAP3}
\Delta_s^{(3)}
=
- u_5^{\frac12} \,
\biggl\{
\frac{u_5^2}{u_1^2+u_2^2+u_3^2+u_4^2}
\,
\biggl[\,\frac{1}{D^2}
\, \Delta_u^{(4)}
-  \frac{1}{D^4}\biggl(2
+ \frac{3}{4}\frac{u_1^2+u_2^2+u_3^2+u_4^2}{u_5^2}\biggr)
\biggr]\,\biggr\} \, u_5^{- \frac12}.
\eea
%=================================================================
The volume elements on ${\mbox S_{4{\bf c}}}$ and ${\mbox S_{3}}$
are given by
%=================================================================
$$
dv(u) = - \frac{D^4}{4}\, {\mbox e^{2i\chi}} \sin\chi
\sin\beta d\chi d\beta d\alpha d\gamma,
\,\,\,\,\,\,\,
dv(s) = R^3 \sin^2\chi\sin\beta d\chi d\beta d\alpha,
$$
%=================================================================
where
%=================================================================
\bea
\frac{u_1^2+u_2^2+u_3^2+u_4^2}{u_5^3}\,
dv(u) = \frac{i}{2D^2}\,
d v(s) \, d\gamma.
\eea
%=================================================================
Integration over $\gamma \in [0, 4\pi]$ gives
%=================================================================
$$
\int_{S_3} .... d\,v(s) =
- \frac{iD^2}{2\pi}
\,
\int_{S_{4C}} .... \frac{u_1^2+u_2^2+u_3^2+u_4^2}{u_5^3} \, d \, v(u).
$$
%=================================================================

\vspace{0.6cm}
\noindent
5.3 {\bf Transformation}  S$_{8C}$$\rightarrow $S$_5$

The Laplace-Beltrami operator on the $u$ - sphere in
($\chi, \vartheta; \alpha, \beta, \gamma, \alpha_H,
\beta_H, \gamma_H$) coordinates is
%=================================================================
$$
\Delta_u^{(8)}
= \frac{2i}{D^2} \sin\chi {\mbox e^{- i\chi}}
\biggl\{\frac{{\mbox e^{- 3i\chi}}}
{\sin^4\chi}\frac{\partial}{\partial\chi}
{\mbox e^{3i\chi}}\sin^4\chi\frac{\partial}{\partial\chi}
+ \frac{1}{\sin^2\chi}\biggl[
\frac{1}{\sin^3\vartheta}\frac{\partial}{\partial\vartheta}
\sin^3\vartheta\frac{\partial}{\partial\vartheta}
$$
\bea
- \frac{4 ({\vec L}^2 + 2{\vec L}\cdot{\vec T}
\sin^2\frac{\vartheta}{2} + {\vec T}^2
\sin^2\frac{\vartheta}{2})}{\sin^2\vartheta}
\biggr]\biggr\},
\eea
%=================================================================
where operator ${\vec L}$ is given by (\ref{L1}) and ${\vec T}$
is
%=================================================================
\bea
\label{T}
{T}_1 &=&
i\left(\cos\alpha_H\cot\beta_H\frac{\partial}{\partial\alpha_H}
+ \sin\alpha_H\frac{\partial}{\partial\beta_H} -
\frac{\cos\alpha_H}{\sin\beta_H}\frac{\partial}{\partial\gamma_H}
\right),
\nonumber\\[2mm]
{T}_2 &=&
i\left(\sin\alpha_H\cot\beta_H\frac{\partial}{\partial\alpha_H}
- \cos\alpha_H\frac{\partial}{\partial\beta_H} -
\frac{\sin\alpha_H}{\sin\beta_H}\frac{\partial}{\partial\gamma_H}
\right),
\\[2mm]
{T}_3 &=&
- i\frac{\partial}{\partial\alpha_H}.
\nonumber
\eea
%=================================================================
The Laplace-Beltrami operator on the five dimensional sphere
in ($\chi, \vartheta; \alpha, \beta, \gamma$) coordinates is
%=================================================================
\bea
\Delta_s^{(5)}
= \frac{1}{R^2}
\left[\frac{1}{\sin^4\chi}\frac{\partial}{\partial\chi}
\sin^4\chi\frac{\partial}{\partial\chi}
+ \frac{1}{\sin^2\chi}\left(
\frac{1}{\sin^3\vartheta}\frac{\partial}{\partial\vartheta}
{\sin^3\vartheta}\frac{\partial}{\partial\vartheta}
- \frac{4 {\vec L}^2}{\sin^2\vartheta}\right)\right].
\eea
%=================================================================
The Laplace-Beltrami operators are related by
%=================================================================
$$
\Delta_u^{(8)}
= \frac{2i}{D^2} \sin\chi {\mbox e^{- \frac{5i}{2}\alpha}}
\biggl[D^4 \Delta_s^{(5)} + \left(\frac94 - 6i\cot\chi\right)
- \frac{1}{\sin^2\chi}
\frac{2{\vec L}\cdot{\vec T} + {\vec T}^2}{\cos^2\frac{\vartheta}{2}}
\biggr]
\, {\mbox e^{\frac{3i}{2}\chi}}
$$
%=================================================================
and  the operator acting on a function of variables
($\chi, \vartheta; \alpha, \beta, \gamma$) is
%=================================================================
\bea
\Delta_s^{(5)} =  - u_9^{\frac32} \,
\biggl\{\frac{1}{D^2} \,
\frac{u_9^2}{\sum_{i=1}^{8}{u_i^2}}
\,
\biggl[\Delta_u^{(8)} - \frac{1}{D^2}\biggl(12
+ \frac{15}{4}\,
\frac{1}{u_9^2}\sum_{i=1}^{8}{u_i^2}
\biggr)\biggr]\,\biggr\} \, u_9^{- \frac32}.
\eea
%=================================================================
The volume elements on S$_{8C}$ and S$_{5}$
have the form
%=================================================================
\bea
dv(u)
&=& - 8 D^8 e^{4i\chi}\, \sin^3\chi
\sin^3\theta d\chi d\theta d\Omega d\Omega_H,
\nonumber\\[2mm]
dv(s)
&=& R^5 \sin^4\chi \sin^3\theta d\chi d\theta d\Omega,
\nonumber
\eea
%=================================================================
where
%=================================================================
\bea
d\Omega = \frac18 \sin\beta d\alpha d\beta d\gamma.
\eea
%=================================================================
We have
%=================================================================
\bea
\frac{1}{u_9^5}
\sum_{i=1}^{8}{u_i^2}\, dv(u)
= \frac{16i}{D^5}\, d v(s) \, d\Omega_H
\eea
%=================================================================
and integration over the variables ($\alpha_H, \beta_H, \gamma_H$)
gives  the formula
%=================================================================
\bea
\label{ORTHOG8}
\int_{S_5} .... \, d\,v(s) =
- \frac{iD^5}{32\pi^2}
\,
\int_{S_{8C}} ....
\sum_{i=1}^{8}{u_i^2}\, \frac{dv(u)}{u_9^5}.
\eea
%=================================================================

\vspace{1cm}
\noindent
{\Large\bf Acknowledgments}

\vspace{0.5cm}
\noindent
We thank Professors A.Odzijewicz, A.N.Sissakian, V.M.Ter-Antonyan
and P.Winternitz, and Drs. A.A.Izmest'ev and L.G.Mardoyan for
interest in this work and very fruitful discussions.Two of the authors E.K. and G.P. thank each others institutions
for kind hospitality during  visits to Dubna and Hamilton.


\begin{thebibliography}{99}
%---------------------------------------------------------------------
\bibitem{LC}
T.Levi-Civita, Sur la Resolution Qualitative di Probleme Restreint des
Trois Corps, {\it Opere Mathematiche}, {\bf 2}, 411-417, 1956.
%----------------------------------------------------------------------
\bibitem{KS}
P.Kustaanheimo and E.Steifel. Perturbation theory of Kepler motion
based on spinor regularisation. {\it J. reine angew Math.}
{\bf 218}, 204, (1965).
%----------------------------------------------------------------------
\bibitem{KIB1}
M.Kibler, T.Negadi.
{\it Croat. Chem. Acta},\ 1{\bf 57(6)},
(1984), 1509-1523.
%----------------------------------------------------------------------
\bibitem{BKM}  C.P.Boyer, E.G.Kalnins, and W.Miller Jr.
St\"ackel-equivalent integrable Hamiltonian systems. {\it SIAM
J. Math. Anal.}, {\bf 17}, 778 (1986).
%-----------------------------------------------------------------------
\bibitem{WT78} W.Thirring. A Course in Mathematical
Physics. 1. Classical Dynamical Systems. E.Harrell, trans., {\it
Springer Verlag}, New York, 1978. (see Section 4.2).
%----------------------------------------------------------------------
\bibitem{WT81} W.Thirring. A Course in Mathematical
Physics. 3. Quantum Mechanics of Atoms and Molecules. E.Harrell, trans.,
{\it Springer Verlag}, New York, 1981. (see Section 4.1).
%----------------------------------------------------------------------
\bibitem{FLUG}
S.Fl\"ugge: {\it Practical Quantum Mechanics}, (Springer-Verlag,
Berlin-Heidelberg-New York, 1971), V1.
%-------------------------------------------------------------------
\bibitem{MAC}
A.Cisneros and H.V.McIntosh. Symmetry of the Two-Dimensional Hydrogen
Atom. {\it J.Math.Phys.} \ {\bf 10}, 277, 1969.
%-------------------------------------------------------------------
\bibitem{MPST}
L.G.Mardoyan, G.S.Pogosyan, A.N.Sissakian and V.M.Ter-Anto\-nyan.
Hidden symmetry, separation of variables and interbasis expansions
in two-di\-men\-sional Hydrogen atom. {\it J.Phys.}\ {\bf A18},
455-466, 1984.
%-------------------------------------------------------------------
\bibitem{IWAI1}
T.Iwai:
Quantization of the conformal Kepler problem and its application to the
hydrogen atom, {\it J.Math.Phys.}, {\bf 23}, 1093, 1982;
The four-dimensional conformal Kepler problem reduces to the
three-dimensional Kepler problem with a centrifugal potential and the
Dirac's monopole field. Classical theory:
{\it J.Math.Phys.}, {\bf 27}, 1523, 1986.
%-----------------------------------------------------------------------
\bibitem{TN}
V.M.Ter-Antonyan and A.Nersessian:
Quantum oscillator and a bound system of two dyons,
{\it Modern Phys.Lett.}, {\bf A10}, 2633-2638, 1995.
%-------------------------------------------------------------------
\bibitem{BP} D.Boccaletti and G.Pucacco. Theory of
orbits. 1. Integrable systems and non pertubative methods.
{\it Springer Verlag}, New York, 1996.
%----------------------------------------------------------------------
\bibitem{DMPST}
L.S.Davtyan, L.G.Mardoyan, G.S.Pogosyan, A.N.Sissakian, and
V.M.Ter-Antonyan;
Generalized KS transformation: from five-dimensional hydrogen atom
to eight-dimensional oscillator,
{\it J.Phys.} {\bf A20}, 6121-6125, 1987.
%----------------------------------------------------------------------
\bibitem{KIB2}
D.Lambert and M.Kibler:
An Algebraic and geometric approach to non-bijective quadratic
transformation, {\it J.Phys.} {\bf A21}, 307, 1988.
%----------------------------------------------------------------------
\bibitem{KIB3}
M.H.Hassan and M.Kibler:
On Hurwitz transformations,
{\it Report LYCEN 9110}, Lyon, 1991
%----------------------------------------------------------------------
\bibitem{HVT}
Le Van Hoang, Tony J.Viloria and Le anh Thu:
On the hydrogen-like atom in five dimensional space:
{\it J.Phys.} {\bf A24}, 3021-3030, 1991.
%----------------------------------------------------------------------
\bibitem{IWAIS}
T.Iwai and T.Sunako:
The quantized SU(2) Kepler problem and its symmetry group
for negative energies,
{\it Journal of Geometry and Physics}, {\bf 20}, 250-272, 1996.
%----------------------------------------------------------------------
\bibitem{MST1}
L.G.Mardoyan, A.N.Sissakian, and V.M.Ter-Antonyan:
8D Oscillator as a hidden SU(2)-monopole,
{\it Particle and Atomic Nuclei}, {\bf 61}, 1746-1750, 1998.
%----------------------------------------------------------------------
\bibitem{PTOL}
M.V.Pletyukhov and E.A.Tolkachev.
8D Oscillator and 5D Kepler problem:
The case of nontrivial
constraints. {\it J.Math.Phys.}, {\bf 40}, 93-100, 1999.
%----------------------------------------------------------------------
\bibitem{SCHR}
E.Schr\"odinger:
A Method of Determining Quantum Mechanical Eigenvalues and Eigenfunctions;
{\it Proc.Roy.Irish Soc.}\ {\bf 46} (1941) 9;
%-------------------------------------------------------------------
Further Studies on Solving Eigenvalue Problems by Factorization;
{\it Proc.Roy.Irish Soc.}\ {\bf 46} (1941) 183;
%-------------------------------------------------------------------
The Factorization of the Hypergeometric Equation;
{\it Proc.Roy.Irish Soc.}\ {\bf 47} (1941) 53
%-------------------------------------------------------------------
\bibitem{HIG}
P.W.Higgs:
Dynamical Symmetries in a Spherical Geometry I;
{\it J.Phys.A: Math.Gen.}\ {\bf 12}, 309, 1979.
%-------------------------------------------------------------------
\bibitem{LEEM}
H.I.Leemon:
Dynamical Symmetries in a Spherical Geometry II;
{\it J.Phys. A: Math. Gen.}\ {\bf 12}, 489, 1979.
%-------------------------------------------------------------------
\bibitem{BOK}
A.A.Bogush, Yu.A.Kurochkin, V.S.Otchik.
{\it DAN BSSR}\ {\bf XXIII}, 19, 1980 (in russian).
%----------------------------------------------------------------------
\bibitem{BIJ1}
A.O.Barut, A.Inomata and G.Junker:
Path integral treatment of the hydrogen atom in a curved
space of constant curvature, {\it J.Phys.} {\bf A20}, 6271-6280,
1987;
Path integral treatment of the hydrogen atom in a curved
space of constant curvature: II. Hyperbolic space.
{\it J.Phys.} {\bf A23}, 1179-1190, 1990;
%----------------------------------------------------------------------
\bibitem{MPSV}
L.G.Mardoyan, G.S.Pogosyan, A.N.Sissakian, S.I.Vinitsky
and T.A.Strizh. A Hydrogen atom in the curved space. Expansion
over free solution. {\it Soviet Journal Nuclear Phys.},
{\bf 56}, 61-73, 1993
%-------------------------------------------------------------------
\bibitem{DAS1}
D.Bonatos, C.Daskaloyannis and K.Kokkotas:
Deformed Oscillator Algebras for Two-Dimensional Quantum Superintegrable
Systems; {\it Phys. Rev.}, {\bf A50}, 3700, 1994.
%-----------------------------------------------------------------------
\bibitem{GRO1}
C.Grosche, G.S.Pogosyan and A.N.Sissakian.
Path Integral discussion for Smorodinsky - Winternitz Potentials:
II. Two - and Three Dimensional Sphere. {\it Fortschritte der
Physik}\ {\bf 43(6)}, 523-563, 1995.
%-------------------------------------------------------------------
\bibitem{KMJP}  E.G.Kalnins, W.Miller Jr. and G.S.Pogosyan.
Superintegrability and associated polynomial solutions. Euclidean space
and the sphere in two dimensions; {\it J.Math.Phys.}\ {\bf 37}, 6439-6467,
1996.
%-----------------------------------------------------------------------
\bibitem{KMJP2}  E.G.Kalnins, W.Miller Jr. and G.S.Pogosyan.
Superintegrability on the two dimensional hyperboloid.
{\it J.Math.Phys.}, {\bf 38}, 5416-5433, 1997.
%-----------------------------------------------------------------------
\bibitem{GRO2}
C.Grosche, G.S.Pogosyan and A.N.Sissakian.
Path Integral Approach to Superintegrable Potentials.
Two - Dimensional Hyperboloid. {\it Phys. Part. Nucl.},
{\bf 27}(3), 244-278, 1996;
Path Integral discussion for Superintegrable Potentials:
IV. Three Dimensional Pseudosphere.
{\it Phys. Part. Nucl.}, {\bf 28}, 486-519, 1997.
%-------------------------------------------------------------------
\bibitem{PS}
G.S.Pogosyan and A.N.Sissakian.
On the Coulomb problem on three dimensional sphere.
{\it Turkish J.Phys.}, {\bf 21}, 515, 1997.
%-------------------------------------------------------------------
\bibitem{IPSW}
A.A.Izmest'ev, G.S.Pogosyan, A.N.Sissakian and P.Winternitz.
Contraction of Lie Algebras and Separation of Variables.
{\it J.Phys.}, {\bf A29}, 5940-5962, 1996.
%-------------------------------------------------------------------
\bibitem{BER}
A.Erdelyi, W.Magnus, F.Oberhettinger and F.Tricomi:
{\it Tables of Integral Transforms}, (McGraw-Hill, New York, 1954),
Vol II.
%-------------------------------------------------------------------
\bibitem{BE}
A.Erdelyi, W.Magnus, F.Oberhettinger and F.Tricomi:
{\it Higher Transdental Functions}, (McGraw-Hill, New York, 1953),
Vols. I and II.
%-----------------------------------------------------------------------
\bibitem{DPST}
L.S.Davtyan, G.S.Pogosyan, A.N.Sissakian and V.M.Ter-Anto\-nyan.
Two-di\-men\-sional Hydrogen atom. Expansion of the polar and
parabolic bases in the continuous spectrum. {\it Teor.Mat.Fiz.},
{\bf 66}, 222-233, 1986.
%-------------------------------------------------------------------
\bibitem{VAR}
D.A.Varshalovich, A.N.Moskalev, and V.K.Khersonskii:
{\it Quantum Theory of Angular Momentum}, (World Scientific, Singapore,
1988).
%-------------------------------------------------------------------
\bibitem{KMT}
Kh.H.Karayan, L.G.Mardoyan and V.M.Ter-Antonyan.
The Eulerian bound states: 5D Coulomb problem.
{\it Preprint JINR, E2-94-359, Dubna, 1994}.
%-------------------------------------------------------------------
\bibitem{HUR}
A.Hurwitz:
Mathematische Werke, Band II, pp.641
({\it Birkh\"auser}, Basel, 1933)
%-----------------------------------------------------------------------

\end{thebibliography}
\end{document}